
\magnification\magstep1
\hfuzz 2pt
\font\bigbf=cmbx10 scaled 1300

\def\go{\mathrel{\raise.3ex\hbox{$>$}\mkern-14mu
             \lower0.6ex\hbox{$\sim$}}}
\def\lo{\mathrel{\raise.3ex\hbox{$<$}\mkern-14mu
             \lower0.6ex\hbox{$\sim$}}}
\def\cI {{\cal I}}
\def\cC {{\cal C}}
\def\cE {{\cal E}}
\def\cF {{\cal F}}
\def\Ib5{{\hbox{\rlap{\hbox{\raise.27ex\hbox{-}}}I}}^{(5)}}
\def\Ibar{{\hbox{\rlap{\hbox{\raise.27ex\hbox{-}}}I}}}
\def\bfe {{\bf e}}
\def\bone {{\bar 1}}
\def\btwo {{\bar 2}}
\def\bthr {{\bar 3}}
\def\bi {{\bar i}}
\def\bj {{\bar j}}
\def\rcm {{r_{cm}}}

\def\bxi{{\vec \xi}}
\def\br{{\bf r}}
\def\xir{{\xi^r}}
\def\xip{{\xi^\perp}}
\def\be{{\bf e}}

\centerline{\bigbf HYDRODYNAMICS OF ROTATING STARS}
\medskip
\centerline{\bigbf AND CLOSE BINARY INTERACTIONS:}
\medskip
\centerline{\bigbf COMPRESSIBLE ELLIPSOID MODELS}

\vskip 0.5truein

\centerline{
DONG LAI$\,$\footnote{$^1$}{Department of Physics, Cornell
University. Email: dong@astrosun.tn.cornell.edu}}
\centerline{\it Center for Radiophysics and Space Research,
Cornell University, Ithaca, NY 14853}

\bigskip
\centerline{
FREDERIC A.~RASIO$\,$\footnote{$^2$}{Hubble Fellow.}}
\centerline{\it Institute for Advanced Study, Olden Lane, Princeton, NJ 08540}

\medskip
\centerline{and}

\smallskip
\centerline{
STUART L.~SHAPIRO$\,$\footnote{$^3$}{Departments of Astronomy and Physics,
Cornell University.}}
\centerline{\it Center for Radiophysics and Space Research,
Cornell University, Ithaca, NY 14853}

\bigskip
\bigskip
\centerline{\bf ABSTRACT}
\medskip

We develop a new formalism to study the dynamics of fluid
polytropes in three dimensions. The stars are modeled as
compressible ellipsoids and the hydrodynamic equations are reduced to
a set of ordinary differential equations for the evolution of the
principal axes and other global quantities.
Both viscous dissipation and the gravitational radiation reaction are
incorporated. We establish the validity of our approximations and demonstrate
the simplicity and power of the method by rederiving
a number of known results concerning the stability
and dynamical oscillations of rapidly rotating polytropes.
In particular, we present a generalization to compressible
fluids of Chandrasekhar's classical results for the secular
and dynamical instabilities of incompressible Maclaurin spheroids.
We also present several applications of our method to astrophysical
problems of great current interest such as the tidal disruption of a star
by a massive black hole, the coalescence of compact binaries driven
by the emission of gravitational waves, and the development of instabilities
in close binary systems.

\bigskip
{\it Subject headings:} celestial mechanics, stellar dynamics --- hydrodynamics
 --- instabilities --- stars: binaries: close --- stars: oscillations
 --- stars: rotation

\vfil\eject
\centerline{\bf 1. INTRODUCTION}
\medskip

In a series of recent papers (Lai, Rasio \& Shapiro 1993a,b, 1994a,b,
hereafter LRS1--4 or collectively LRS), we have developed a
new analytic method to calculate ellipsoidal figures of
equilibrium both for a single rotating polytrope and for polytropes in binary
systems. Our approach is based on the use of an energy variational
principle to construct approximate equilibrium solutions and to
determine their dynamical and secular stability.
In addition to providing compressible generalizations
for all the classical incompressible solutions (Chandrasekhar 1969,
hereafter Ch69), our method can also be applied to more
general models of binary systems where the stellar masses, radii, spins,
entropies, and polytropic indices are all allowed to vary
independently for each component. As a result, we were able to study
the equilibrium and
stability properties of many different types of fairly realistic binary models
(see especially LRS4).

We have used the method to explore the implications of instabilities
for the  coalescence of close binary systems (LRS2, LRS3). Of particular
importance currently are
coalescing neutron star binaries, which are the primary targets for the
detection of gravitational waves by LIGO (Abramovici et al.\ 1992; LRS3).
In our previous papers we have studied these systems
by considering  equilibrium sequences
of binary configurations, treating the binary separation $r$ as a
dynamical variable but with all other variables assuming their
equilibrium values.  This is adequate to capture the essential features
of the evolution, but does not provide quantitatively accurate results
when the orbital evolution takes place on a timescale comparable to the
internal hydrodynamic timescale.
The purpose of the present paper is to extend our formalism and
develop a fully dynamical theory for the evolution of compressible
ellipsoids.

Detailed studies of hydrodynamic interactions between stars normally require
large-scale numerical simulations in three dimensions (see, e.g., Lai, Rasio,
\& Shapiro 1993c; Rasio \& Shapiro 1991, 1994). These simulations demand
extensive
computational resources and cannot be used to cover a large parameter space.
In our treatment, we replace the infinite number of
degrees of freedom in a fluid system
by a small number of variables specifying the essential geometric
and kinematic properties of the system.
The dynamics is then described approximately
by a set of ordinary differential equations (ODEs) for the time evolution of
these variables. This represents an enormous simplification. However,
for many problems, useful insights and even reliable quantitative
results can be obtained using such an approach. The simplicity of
a formulation in terms of ODEs allows for extensive coverage of
the parameter space. In addition, we can
follow the evolution of a system over a large number
of dynamical times without having to worry about excessive computational
time or about the growth of numerical errors. This advantage  is crucial
for studying the secular evolution of a system on a dissipative timescale
while still allowing for dynamical processes. A distinctive example
is the coalescence of binary neutron stars, which begins as a very slow
orbital decay driven by the emission of gravitational waves, but
ends in a rapid hydrodynamic merging of the two stars after the stability
limit has been reached (LRS3; Rasio \& Shapiro 1994).

In the incompressible limit, the dynamics of an isolated
ellipsoid was first formulated by Lebovitz (1966; see also Ch69, Chap.~4).
This so-called Riemann-Lebovitz formulation of the problem forms the
basis of many subsequent studies and astrophysical applications of
ellipsoidal models. In particular, the effects of fluid viscosity and
gravitational radiation reaction on the stability of a rapidly rotating star
were investigated by Chandrasekhar (1970), Press \& Teukolsky (1973), Miller
(1974), and Detweiler \& Lindblom (1977). The Riemann-Lebovitz equations
have also been used to study the gravitational collapse of rotating
stars and the resulting gravitational radiation by
relaxing the assumption of incompressibility
and incorporating a polytropic equation of state while still assuming a
homogeneous density profile (see Shapiro 1979 and
references therein). Nduka (1971) has studied the dynamics of an
incompressible ellipsoid in a parabolic orbit around a fixed point mass.

For compressible systems, Carter \& Luminet (1983, 1986)
have developed a dynamical theory
of an affine stellar model in the context of tidal interactions with
a massive black hole. Kochanek (1992a,b) has used the method
to study the tidal capture of a polytrope by a point mass and the
coalescence of two polytropes in a close binary.
In the affine model, the fluid compressibility is treated under the
assumption that the surfaces of constant density remain self-similar
ellipsoids. Our dynamical model for compressible ellipsoids
is essentially equivalent to the affine model (cf.\ LRS1),
but our formulation of the problem is quite different and our dynamical
equations more closely resemble the Riemann-Lebovitz equations in the
incompressible limit. In contrast to previous studies,
we present a completely general formulation for isolated stars
and for a star on a bound or unbound trajectory around a point mass.
We incorporate both viscosity and gravitational radiation reaction
as possible dissipation mechanisms.
Our formulation makes explicit use of global quantities
such as the total angular momentum and fluid circulation which are conserved
in the absence of dissipation. This greatly simplifies
the description of many dynamical processes, and we feel that this
simplification has not been fully appreciated in previous studies
based on the affine model.

The main purpose of this paper is to formulate general  dynamical equations
for compressible ellipsoids and to present a small survey of astrophysical
applications. We wish to illustrate how these general equations
can be used to tackle a variety of multidimensional hydrodynamic
problems of great current interest.
The dynamical equations for an isolated ellipsoid supported
by a polytropic equation of state are derived
in \S 2, where our general assumptions are also presented (\S 2.1).
Many of the expressions derived in \S 2 can be readily
transported to more general situations. In \S 3 we consider the dynamical
stability and oscillations of a single rotating polytrope.
In \S 4, we incorporate viscous dissipation and gravitational
radiation reaction into the dynamical equations,
and we study the secular evolution of isolated rotating stars.
In \S 5, the dynamical equations derived in \S 2 for single stars
are generalized to a binary system (either bound or unbound)
consisting of a polytrope and a point mass.
The dynamical stability of general Roche-Riemann binary models is considered
in \S 6. In \S 7, we study tidal encounters of a star with a massive body,
and we compare our results with those of linear perturbation theory as well
as previous nonlinear calculations. In \S 8, we consider the evolution of
binary systems driven by viscous dissipation. In \S 9, we incorporate
gravitational radiation reaction in our treatment of binaries, and we
study the orbital evolution driven by gravitational radiation.

\vfill\eject
\bigskip
\centerline{\bf 2. DYNAMICAL EQUATIONS FOR SINGLE STARS:}
\smallskip
\centerline{\bf RIEMANN-S ELLIPSOIDS}
\medskip

In this section, we derive the equations of motion for an isolated
compressible Riemann-S ellipsoid. Our equations represent a generalization
of the Riemann-Lebovitz equations, which only apply to an incompressible fluid
(Ch69, Chap.~4).

\bigskip
\centerline{\bf 2.1 Assumptions}
\medskip

Throughout this paper, we adopt a polytropic equation of state for
the fluid,
$$P=K\rho^{1+1/n}.\eqno(2.1)$$
Our basic assumptions concerning the interior structure of the star can be
summarized as follows (see LRS1 for more details).
We assume that the surfaces of constant density
can be represented approximately by {\it self-similar ellipsoids\/}.
The geometry is then
completely specified by the three principle axes of the outer surface.
Furthermore, we assume that the density profile $\rho(m)$ inside each star,
where $m$ is the mass interior to an isodensity surface,
is identical to that of a {\it spherical\/} polytrope with the same volume.
In the reference frame comoving with the star's center of mass,
the velocity field of the fluid is modeled as a linear superposition of
three components: (1) a rigid rotation of the ellipsoidal
{\it figure\/}; (2) an internal fluid circulation with {\it uniform
vorticity\/}; and
(3) an ellipsoidal expansion or contraction (see eqs.~[2.3] and~[2.4] below).

Under these assumptions, the number of internal degrees of freedom
for each star is reduced to nine in general:
the three principal axes $a_1$, $a_2$, $a_3$
of the outer surface, the three components of the angular velocity
of the ellipsoidal figure $\Omega_1$, $\Omega_2$, $\Omega_3$,
and the three components of the vorticity $\zeta_1$, $\zeta_2$, $\zeta_3$.
We shall further restrict ourselves to the cases where
both the angular velocity  and the vorticity  are parallel to one
of the principal axes, here taken to be the $z$ or $a_3$ axis by convention.
This choice corresponds to the Riemann ellipsoids of type-S
(Ch69). As we show in \S2.2, under these conditions,
the exact hydrodynamic equations
(the Euler and Navier-Stokes equations, with and without gravitational
radiation reaction) are replaced by a set of ODEs
for the time evolution of the principle axes $a_1$, $a_2$, $a_3$,
the angular velocity of the ellipsoidal figure $\Omega=\Omega_3$,
 and the angular frequency
of the internal circulation $\Lambda\propto\zeta_3$ (cf.\ LRS1, \S5.1).

As we noted in the introduction, our model is formally
equivalent to the affine model developed extensively by
Carter and Luminet (1985; Luminet \& Carter 1986).
In the incompressible limit ($n=0$), both models can be obtained
equivalently by imposing fixed holonomic constraints on the system,
requiring the fluid surface to be ellipsoidal and the fluid velocity to be a
linear function of coordinates. This is known as the Dirichlet problem
(see Ch69, Chap.~4).
In the $n=0$ limit, our dynamical equations reduce explicitly
to the equations obtained in the Riemann-Lebovitz
formulation of the Dirichlet problem (Ch69, \S27).
For a single isolated star with $n=0$, the solutions
we derive represent {\it exact\/}
solutions of the true hydrodynamic equations. For $n\ne0$, our solutions
are only approximate since the isodensity surfaces can no longer be exactly
ellipsoidal, and the velocity field of the fluid cannot be described
exactly by a linear function of coordinates (see Ipser \& Managan 1981 for
a formal proof). For binary systems, our solutions are always approximate
because we truncate the gravitational interaction to quadrupole order
(see \S5 below).

Clearly, our formalism cannot be applied to hydrodynamic processes such
as stellar collisions that are violent enough to disrupt a star completely
or to produce significant shock heating.
Even for dynamical processes that are not very disruptive,
our treatment allows us to follow only a small subset of the
the many degrees of freedom that may be important. This is especially
true for compressible systems, which have typically many more important
degrees of freedom than incompressible systems.
Consider for example small nonradial oscillations
of a star (see, e.g., Cox 1980). For $n=0$, only the f-modes of oscillation
 exist. But when $n\ne0$, there are many additional
p-modes and possibly also g-modes of oscillation,
corresponding to nonellipsoidal disturbances of the star.
Such nonellipsoidal motions cannot be described at all within our simplified
treatment. Even for $n=0$, an ellipsoidal model can only represent the
$l=2$, quadrupolar f-modes of oscillation. These modes often dominate the
response of a star to tidal perturbations by passing objects
(see Kochanek 1992a and \S7 below), but they are not sufficient for a
complete description of the dynamical response (cf.\ Press \& Teukolsky 1977).

\bigskip
\centerline{\bf 2.2 Derivation of The Dynamical Equations}
\medskip

Let ${\bf e}_1$, ${\bf e}_2$, and ${\bf e}_3$ be the basis unit vectors
along the instantaneous directions of the principal axes of the
ellipsoid, with ${\bf e}_3$ along the rotation axis
(we refer to this as the ``{\it body frame}''). In the inertial
frame, the velocity field ${\bf u}$ inside the ellipsoid can be written as
$${\bf u}={\bf u}_s+{\bf u}_e.\eqno(2.2)$$
Here ${\bf u}_s$ is the velocity field of an equilibrium Riemann-S ellipsoid,
$${\bf u}_s=\left({a_1\over a_2}\Lambda-\Omega\right)x_2{\bf e}_1
+\left(-{a_2\over a_1}\Lambda+\Omega\right)x_1{\bf e}_2,\eqno(2.3)$$
where $\Omega$ is the angular velocity of the ellipsoidal figure, and
 $\Lambda$ is the angular frequency of the internal fluid circulation
(cf.\ LRS1, \S5.1). For nonequilibrium configurations, we add
the velocity ${\bf u}_e$ describing an ellipsoidal expansion or contraction
of the star,
$${\bf u}_e={\dot a_1\over a_1}x_1{\bf e}_1+{\dot a_2\over a_2}x_2{\bf e}_2+
{\dot a_3\over a_3}x_3{\bf e}_3.\eqno(2.4)$$
The kinetic energy is then given by
$$T=\int\!d^3x\,{1\over 2}\rho\, {\bf u}^2=T_s+T_e.\eqno(2.5)$$
Here $T_s$ is the ``spin'' kinetic energy (rotation and internal circulation
of the fluid; cf.\ LRS1, eq.~[5.6]),
$$T_s={1\over2}I(\Lambda^2+\Omega^2)
-{2\over5}\kappa_nMa_1a_2\Lambda\Omega,\eqno(2.6)$$
where $I$ is the moment of inertia,
$$I={1\over5}\kappa_nM(a_1^2+a_2^2),\eqno(2.7)$$
$\kappa_n$ is a structure constant depending on the polytropic index
(numerical values are given in LRS1, Table~1),
and $T_e$ is the kinetic energy associated with the expansion or
contraction,
$$T_e={1\over 10}\kappa_nM(\dot a_1^2+\dot a_2^2+\dot a_3^3).\eqno(2.8)$$

The total internal energy $U$ of the fluid is given by
$$U=\int\! {nP\over \rho}\, dm = k_1K\rho_c^{1/n}M,\eqno(2.9)$$
where $k_1$ is a constant depending on the polytropic index, and
$\rho_c$ is the central density, equal to that of a spherical polytrope
with the same mass and volume in our approximation (see LRS1, \S2.1).
The self-gravitational potential energy is given by
$$W=-{3\over5-n}{GM^2\over R}f,\eqno(2.10)$$
where $R\equiv (a_1a_2a_3)^{1/3}$ is the mean radius of the ellipsoid, and
$$f={\cI\over 2R^2},~~~~{\rm with}~~~~
\cI=A_1a_1^2+A_2a_2^2+A_3a_3^2\eqno(2.11)$$
($f=1$ for spherical configurations).
The dimensionless index symbols $A_i$ are defined as in Ch69 (\S17).

The Lagrangian governing the dynamics of the ellipsoid can now be written as
$$L(q_i;\dot q_i)=T-U-W,\eqno(2.12)$$
where $\{q_i\}=\{a_1,a_2,a_3,\phi,\psi\}$, and
$\{\dot q_i\}=\{\dot a_1,\dot a_2,\dot a_3,\Omega,\Lambda\}$ (we have
introduced angles $\phi$ and $\psi$ such that $\dot \phi=\Omega$ and
$\dot\psi=\Lambda$). The dynamical equations are obtained from the
Euler-Lagrange equations
$${d\over dt}{\partial L\over\partial \dot q_i}=
{\partial L\over\partial q_i}.\eqno(2.13)$$
Using the relation (Ch69, Chap.~2)
$${\partial \cI\over \partial a_i}={1\over a_i}(\cI-a_i^2A_i),\eqno(2.14)$$
equation (2.13) for $q_i=a_1$ can be written as
$${1\over 5}\kappa_nM\ddot a_1={1\over 5}\kappa_n Ma_1(\Omega^2+\Lambda^2)
-{2\over 5}\kappa_nMa_2\Omega\Lambda+{k_1M\over na_1}{P_c\over\rho_c}
-{2\pi\over 5-n}MG {\bar\rho} a_1A_1,\eqno(2.15)$$
where $\bar\rho=3M/(4\pi R^3)$ is the mean density. Equations for
$a_2$ and $a_3$ can be similarly obtained.
For $q_i=\phi$ and $\psi$, equation (2.13) yields the conservation laws for
angular momentum and circulation,
$${dJ_s\over dt}=0,~~~~{d\cC\over dt}=0,\eqno(2.16)$$
where $J_s$ is the angular momentum,
$$J_s ={\partial L\over \partial \Omega}=
I\Omega -{2\over5}\kappa_nMa_1a_2\Lambda,\eqno(2.17)$$
and $\cC$ is the fluid circulation (cf. LRS1)
$$\cC ={\partial L\over \partial\Lambda}=
I\Lambda-{2\over5}\kappa_nMa_1a_2\Omega.\eqno(2.18)$$

The complete set of dynamical equations can be rewritten in the form
$$\eqalignno{
&\ddot a_1 =a_1(\Omega^2+\Lambda^2)-2a_2\Omega\Lambda
-{2\pi G\over q_n}a_1A_1\bar\rho
+\left({5k_1 \over n\kappa_n}\right){P_c\over\rho_c}{1\over a_1},  &(2.19)\cr
&\ddot a_2 =a_2(\Omega^2+\Lambda^2)-2a_1\Omega\Lambda
-{2\pi G\over q_n}a_2A_2\bar\rho
+\left({5k_1 \over n\kappa_n}\right){P_c\over\rho_c}{1\over a_2},  &(2.20)\cr
&\ddot a_3 =-{2\pi G\over q_n}a_3A_3\bar\rho
+\left({5k_1 \over n\kappa_n}\right){P_c\over\rho_c}{1\over a_3},  &(2.21)\cr
&{d\over dt}\left(a_1\Omega-a_2\Lambda\right)
=-\dot a_1\Omega+\dot a_2\Lambda,				    &(2.22)\cr
&{d\over dt}\left(-a_2\Omega+a_1\Lambda\right)
=\dot a_2\Omega-\dot a_1\Lambda,			 	    &(2.23)\cr
}$$
where $q_n\equiv \kappa_n(1-n/5)$. Rather than using equations~(2.22)
and~(2.23) as such, we shall use the equivalent pair
$$\eqalignno{
\dot\Omega &=\left({a_2\over a_1}-{a_1\over a_2}\right)^{-1}
\left[2\left({\Omega\over a_2}+{\Lambda\over a_1}\right)\dot a_1
-2\left({\Omega\over a_1}+{\Lambda\over a_2}\right)\dot a_2
\right],      \qquad&(2.24)\cr
\dot\Lambda &=\left({a_2\over a_1}-{a_1\over a_2}\right)^{-1}
\left[2\left({\Omega\over a_1}+{\Lambda\over a_2}\right)\dot a_1
-2\left({\Omega\over a_2}+{\Lambda\over a_1}\right)\dot a_2
\right].     &(2.25)}$$

In addition to the conservation of total angular momentum and circulation
(eqs.~[2.17]-[2.18]),
it is easy to check that the total energy is also conserved,
i.e.,
$$\cE=\cE_s=T+U+W={\rm constant}.\eqno(2.26)$$
For ${\ddot a}_i=0$, one can verify that
equations (2.19)--(2.21) reduce to the equilibrium equations
for Riemann-S ellipsoids derived in LRS1 (\S5.1).

\vfill\eject
\bigskip
\centerline{\bf 2.3 The Pressure Term}
\medskip

To implement the dynamical equations in numerical integrations,
it is convenient to express the
pressure term $(5k_1P_c)/(n\kappa_n\rho_c)$ in terms of $R_o$, $M$ and
other dynamical variables, where
$R_o$ is the radius of a nonrotating spherical polytrope
of the same mass $M$ and polytropic constant $K$. This is done as follows.
Since $P_c/\rho_c=K\rho_c^{1/n}\propto R^{-3/n}$, we can write
$${5k_1 \over n\kappa_n}{P_c\over\rho_c}=C_nR^{-3/n},\eqno(2.27)$$
where $C_n$ is a constant depending only on $n$, $K$, $M$. We can
obtain the value of $C_n$ by considering a single, nonrotating spherical
polytrope in equilibrium. For such a configuration, equations (2.19)--(2.21)
become
$$0=-{2\pi G\over q_n}R_oA_1\bar\rho_o+C_nR_o^{-3/n}{1\over R_o}.
\eqno(2.28)$$
As $A_1=2/3$ for a sphere, we get
$$C_n={GM\over q_n}R_o^{3/n-1}.\eqno(2.29)$$
Therefore, in equations (2.19)--(2.21), we can use
$${5k_1 \over n\kappa_n}{P_c\over\rho_c}
={GM\over q_nR_o}\left({R_o\over R}\right)^{3/n},~~~~~~(n\ne 0).\eqno(2.30)$$

Expression~(2.30) is obviously not valid in
the incompressible case. For $n\rightarrow 0$,
$k_1/n\rightarrow 2/5$ and $\kappa_n\rightarrow 1$, so that
$${5k_1 \over n\kappa_n}{P_c\over\rho_c}={2P_c\over\rho_c}.~~~~~~
(n=0)\eqno(2.31)$$
To evaluate $2P_c/\rho_c$ in this limit, we need to use the incompressible
condition
$$a_1a_2a_3=R_o^3={\rm constant},\eqno(2.32)$$
which gives
$${\dot a_1\over a_1}+{\dot a_2\over a_2}+{\dot a_3\over a_3}=0.\eqno(2.33)$$
Now adding (2.19)$/a_1$, (2.20)$/a_2$ and (2.21)$/a_3$,
and using the above expression, we get
$$\eqalign{
{5k_1 \over n\kappa_n}{P_c\over\rho_c}&={2P_c\over\rho_c} \cr
 &=\left(\sum_i\!\!{1\over a_i^2}\right)^{-1}
 \left[\sum_i\!\!\left({\dot a_i\over a_i}\right)^2-2(\Omega^2+\Lambda^2)
 +2\Omega\Lambda\left({a_2\over a_1}+{a_1\over a_2}\right)
 +4\pi G\bar\rho\right],~~~~(n=0)
}\eqno(2.34)$$
where we have used $A_1+A_2+A_3=2$. From equation (2.30), we see that
the internal energy (2.9) can also be written as
$$U=k_1{P_c\over \rho_c}M={n\over 5-n}{GM^2\over R_o}
\left({R_o\over R}\right)^{3/n}.\eqno(2.35)$$
Note that $U=0$ for $n=0$ since $k_1=0$ (cf.\ LRS1, eq.~[2.9]).

\bigskip
\centerline{\bf 3. DYNAMICAL OSCILLATIONS AND STABILITY OF SINGLE STARS}
\medskip

In this section, we use the dynamical equations derived in \S 2 to study
the stability and dynamical oscillations of Riemann-S ellipsoids.
We first establish a general criterion for linear stability
to  small-amplitude dynamical perturbations (\S 3.1).
We then study the dynamical oscillations of an  ellipsoid, including
finite-amplitude self-similar pulsations (\S 3.2) and small-amplitude
nonradial oscillations (\S 3.3). Most of the results obtained in this section
are well-established. The derivations presented here will demonstrate that
our formalism can reproduce these results or generalize them to the
compressible case.

\bigskip
\centerline{\bf 3.1 Dynamical Stability Criterion}
\medskip

Here we show explicitly that the onset of dynamical instability
along an equilibrium sequence
can be determined from an appropriate energy function defined along that
sequence, as we have done in our previous papers (LRS).
To do so we only need to recast the
dynamical equations into Hamiltonian form. From the Lagrangian
(2.12), the canonical momenta associated with the
generalized coordinates $\{q_i\}=\{a_1,a_2,a_3,\phi,\psi\}$ are calculated
as $P_i={\partial L/\partial q_i}$, giving
$$P_{a_i}={1\over 5}\kappa_nM\dot
a_i,~~~P_{\phi}=J_s,~~~P_{\psi}=\cC.\eqno(3.1)$$
The Hamiltonian is then
$$\eqalign{
H &= \sum_i\dot q_iP_i-L \cr
  &= {1\over 2(\kappa_nM/5)}(P_{a_1}^2+P_{a_2}^2+P_{a_3}^2)
+E(a_1,a_2,a_3;J_s,\cC).
}\eqno(3.2)$$
The energy function $E$ plays the
role of an effective potential, and is given by (cf.\ LRS4,
eqs.~[2.23] and~[2.24], where we make explicit the conserved
quantities $J_s$ and $\cC$)
$$E(a_1,a_2,a_3;J_s,\cC)={1\over 2I_{+}}(J_s+\cC)^2
+{1\over 2I_{-}}(J_s-\cC)^2+U+W,\eqno(3.3)$$
where
$$I_{\pm}\equiv {2\over 5}\kappa_nM(a_1\mp a_2)^2.\eqno(3.4)$$
Hamilton's equations can then be written as
$${1\over 5}\kappa_nM\ddot a_i=-{\partial E\over \partial a_i},\eqno(3.5)$$
together with $\dot J_s=\dot\cC=0$.
For linear perturbations we write
$$a_i=a_{i,eq}(1+\alpha_i),\eqno(3.6)$$
where $|\alpha_i|\ll1$, and equation (3.5) becomes
$$I_{ii}\ddot\alpha_i=-\sum_j\left[\left({\partial^2 E\over \partial
a_i\partial a_j}\right)a_ia_j\right]_{eq}\alpha_j~~~~({\rm no~sum~over}~i),
\eqno(3.7)$$
where $I_{ij}=(\kappa_n/5)Ma_i^2\,\delta_{ij}$.
If we let $\alpha_i \propto e^{i\sigma t}$, equation (3.7) gives
$$\sigma^2=0 \Longleftrightarrow {\rm det}\left({\partial^2 E\over \partial
a_i\partial a_j}\right)_{eq}=0.\eqno(3.8)$$
Here a subscript ``eq'' means evaluating the function for equilibrium
configurations. Equation~(3.8) is the condition that determines the onset of
instability along a sequence of equilibrium configurations (see LRS1, \S2).

\bigskip
\centerline{\bf 3.2 Homologous Pulsations}
\medskip

Consider the homologous perturbations described by
$$a_i(t)=\xi(t)a_{i,eq},\eqno(3.9)$$
about a compressible Riemann-S equilibrium configuration. Here
$\xi(t)$ is not necessarily close to unity (i.e., we allow for
finite-amplitude oscillations). Such oscillations only exist for
a compressible ($n\neq 0$) configuration. Our description of the
oscillations by equation~(3.9) is only approximate (cf.~\S 3.3),
but it is a good approximation  for slowly rotating configurations.

Adding $(\kappa_nMa_1/5)\times\,$eq.\ (2.19),
$(\kappa_nMa_2/5)\times\,$eq.\ (2.20), and
$(\kappa_nMa_3/5)\times\,$eq.\ (2.21) we obtain
$$I_{t,eq} \xi\ddot\xi=2 T_s-{|W_{eq}|\over \xi}
+\left({3k_1MP_c\over n\rho_c}\right)_{eq}\xi^{-3/n},\eqno(3.10)$$
where $I_t\equiv I_{11}+I_{22}+I_{33}$, $T_s$ is given by equation (2.6),
and we have used equation (2.10) and the fact that
$P_c/\rho_c\propto (a_1a_2a_3)^{-1/n}$. Since $T_s$ can be written as
(LRS4, eqs.~[2.23]--[2.24])
$$T_s={1\over 2I_{+}}(J_s+\cC)^2+{1\over 2I_{-}}(J_s-\cC)^2,\eqno(3.11)$$
and $J_s,\,\cC$ remain  constant during a dynamical perturbation, we have
$$T_s=T_{s,eq}\xi^{-2}.\eqno(3.12)$$
For the last term in equation (3.10), we use equation (2.30) and the expression
$$\left({R\over R_o}\right)_{eq}=\left[f\left(1-2{T_s\over |W|}\right)\right]
^{-n/(3-n)},\eqno(3.13)$$
(LRS1, eq.\ [4.25]). Equation (3.10) then reduces finally to
$$I_t\ddot
\xi=|W|(\xi^{-1-3/n}-\xi^{-2})+2T_s(\xi^{-3}-\xi^{-1-3/n}),\eqno(3.14)$$
where we have suppressed the subscript ``eq''.
This equation was first derived for an axisymmetric, uniformly
rotating polytrope
by Tassoul (1970). We see that it applies also to our more general
compressible Riemann-S ellipsoids.

Consider now small-amplitude oscillations, with
$\xi=1+\xi_1$ and $|\xi_1|\ll1$.
Linearization of equation~(3.14) yields
$$I_t\ddot\xi_1=|W|\left[(4-3\Gamma)-{2T_s\over |W|}
(5-3\Gamma)\right]\xi_1.\eqno(3.15)$$
We see that stability requires
$$\Gamma>\Gamma_{crit}={4\over 3}-{2T_s/|W|\over 3(1-2T_s/|W|)}.\eqno(3.16)$$
The corresponding oscillation frequency is given by
$$\sigma^2={|W|\over I_t}\left[(3\Gamma-4)
-2{T_s\over |W|}(3\Gamma-5)\right].\eqno(3.17)$$
This is a well-known result (the Ledoux formula), first derived for uniformly
and slowly rotating stars (see Tassoul 1978, \S14.2). We see that it can also
be applied (approximately) to more general ellipsoidal configurations, even
when the rotation rate is large.

For nonrotating stars, equation (3.17) gives the
fundamental pulsation mode frequency
$$\sigma_o^2={|W|\over I_t}(3\Gamma-4)={GM\over R_o^3}{1\over q_n}
(3\Gamma-4),\eqno(3.18)$$
where $q_n=(1-n/5)\kappa_n$, and $\kappa_n$ is defined by equation~(2.7).
The values of $\kappa_n$ for different $n$ are given in Table~1 of
LRS1. In Table~1 here, we compare the predictions of equation~(3.18)
with the exact results obtained by integrating numerically
the radial pulsation equation for polytropes with
$\Gamma=\Gamma_1$ (Cox 1980, Chap.~8).
We see that equation~(3.18) predicts the exact results remarkably well.
The discrepancy remains less than $\sim 3\%$ for all values of $n$.
For $n=3$ and in the limit as $n\rightarrow 0$,
equation~(3.18) yields the exact results. This is not surprising
since, in these limits, small radial pulsations in the fundamental
mode  are indeed homologous.
The excellent agreement for all $n$ gives us confidence that our treatment of
small dynamical perturbations is always a very good approximation.

We can also consider the stability to finite-amplitude oscillations
(Tassoul 1970).
Integrating equation~(3.14) once, we obtain
$${1\over 2}I_t\dot\xi^2+|W|\Psi(\xi)={\rm constant},\eqno(3.19)$$
where
$$\Psi(\xi)={T_s/|W|-\xi\over\xi^2}+{1-2T_s/|W|\over
3(\Gamma-1)\xi^{3(\Gamma-1)}}.\eqno(3.20)$$
{}From equation~(3.19), we see that $\Psi(\xi)$ acts like an effective
potential for the oscillation.
It is easy to see that $\xi=1$ is a local minimum of $\Psi(\xi)$ only when
the condition~(3.16) is satisfied. But even then it is not necessarily a
global minimum. Consider the behavior of $\Psi(\xi)$ for $\xi\gg1$,
$$\Psi(\xi)\simeq {1\over \xi}\left[-1+{1-2T_s/|W|\over
3(\Gamma-1)\xi^{3\Gamma-4}}\right].\eqno(3.21)$$
We see that when $\Gamma<4/3$, $\Psi(\xi)$ has a global minimum at
$\xi=\infty$.
Thus for $\Gamma_{crit}<\Gamma<4/3$, the star is only metastable (Tassoul
1970).
The condition for stability to finite-amplitude dynamical perturbations is
$\Gamma>4/3$, independent of rotation.

\bigskip
\centerline{\bf 3.3 General Oscillations of a Maclaurin Spheroid}
\medskip

We now consider the general linear oscillations
$$a_i(t)=a_{i,eq}(1+\alpha_i),\eqno(3.22)$$
with $|\alpha_i|\ll1$. For simplicity we assume $a_{1,eq}=a_{2,eq}$,
i.e., the equilibrium configuration is a compressible Maclaurin spheroid.

We must linearize equations (2.19)--(2.23) to obtain the dynamical
equations for
$\alpha_i$. In the derivation, one must be careful to note that
a Maclaurin spheroid need not be assigned the angular velocity of the frame
in which it is at rest, i.e., the value of $\Omega_{eq}$ and $\Lambda_{eq}$
must be determined, like the frequency $\sigma$, from the analysis
(Rossner 1967). Here we will use equation~(3.7) directly.
For a Maclaurin spheroid we have $J_s=-\cC$
(cf.\ eqs.~[2.17]--[2.18] for $a_1=a_2$), and the kinetic energy term is simply
$$T_s={J_s^2\over (\kappa_n/5)M(a_1+a_2)^2}.\eqno(3.23)$$
Setting the first derivatives of $E$ equal to zero,
we obtain the equilibrium conditions
$$\eqalign{
T_s &=-\Sigma (a_1^2A_1-a_3^2A_3),\cr
{1\over n}U&=-\Sigma a_3^2A_3,
}\eqno(3.24)$$
where we have defined
$$\Sigma\equiv -{3GM^2\over 2(5-n)R^3},\eqno(3.25)$$
so that $W=\Sigma\cI$. For convenience, we also define
$$\cE_{ij}\equiv \left(a_ia_j{\partial^2E\over\partial a_i\partial
a_j}\right)_{eq}.\eqno(3.26)$$
After some algebra we obtain the following expressions,
$$\eqalign{
\cE_{11} &=\cE_{22}=
{1\over n}\left(1+{1\over n}\right)U-a_1^2\Sigma(3B_{11}-2A_1)+{3\over
2}T_s,\cr
\cE_{33} &=
{1\over n}\left(1+{1\over n}\right)U+2a_3^2\Sigma B_{13},\cr
\cE_{12} &=
{1\over n^2}U-a_1^2\Sigma(B_{11}-A_1)+{3\over 2}T_s,\cr
\cE_{13} &=\cE_{23}={1\over n^2}U-a_1^2\Sigma (B_{13}-A_1),
}\eqno(3.27)$$
where the $B_{ij}$ are defined as in Ch69 (Chap.~3) and the right-hand sides
are evaluated for the equilibrium configuration (the subscript ``eq''
has been dropped).
Let $\alpha_i \propto e^{i\sigma t}$. The linearized equations for small
oscillations
then give
$$\left(\matrix{\cE_{11}-I_{11}\sigma^2 &\cE_{12} &\cE_{13}\cr
		\cE_{12} &\cE_{11}-I_{11}\sigma^2 &\cE_{13}\cr
		\cE_{13} &\cE_{13} &\cE_{33}-I_{33}\sigma^2\cr}\right)
  \left(\matrix{\alpha_1\cr
		\alpha_2\cr
		\alpha_3\cr}\right)=0.
\eqno(3.28)$$

We now determine the eigenfrequencies and eigenmodes by solving the linear
system~(3.28). One of the eigenvalues can be obtained from
$$I_{11}\sigma^2=\cE_{11}-\cE_{12}={1\over
n}U-a_1^2\Sigma(2B_{11}-A_1).\eqno(3.29)$$
Using the equilibrium conditions~(3.24), we obtain
$$\bar\sigma^2={4\over q_n}B_{11}-\bar\Omega^2,\eqno(3.30)$$
where $\bar\sigma\equiv\sigma/(\pi G\bar\rho)^{1/2}$,
$\bar\Omega\equiv\Omega/(\pi G\bar\rho)^{1/2}$,
and $\bar\rho=3M/(4\pi R^3)$ is the mean
density. The corresponding eigenmode has the form
$$\left(\matrix{\alpha_1\cr
                \alpha_2\cr
                \alpha_3\cr}\right)\propto \left(\matrix{1\cr
			                   -             1\cr
        			                         0\cr}\right).
\eqno(3.31)$$
This mode is the compressible generalization of the {\it toroidal mode\/}
found in Ch69 (or bar mode).
In a reference frame corotating with the unperturbed star,
fluid elements oscillate at a frequency
$\sigma_o$ given by
$$\bar\sigma_o=\bar\Omega\pm\bar\sigma
=\bar\Omega\pm\left({4\over q_n}B_{11}-\bar\Omega^2\right)^{1/2}.\eqno(3.32)$$
This expression agrees with the result~(5.50) of Ch69 for $n=0$.
We see that the onset of dynamical instability is given by
$$\bar\Omega^2=\bar\Omega^2_{dyn}\equiv{4\over q_n}B_{11}.\eqno(3.33)$$
The toroidal mode is neutrally stable as seen in the corotating frame
($\sigma_o=0$) when
$$\bar\Omega^2=\bar\Omega^2_{bif}\equiv{2\over q_n}B_{11},\eqno(3.34)$$
corresponding to the point where the Jacobi sequence bifurcates from
the Maclaurin sequence (Ch69; LRS1).

The eigenfrequencies for the two other modes are determined from
$$(\cE_{11}+\cE_{12}-I_{11}\sigma^2)(\cE_{33}-I_{33}\sigma^2)=2\cE_{13}^2,
\eqno(3.35)$$
and the corresponding eigenmodes have the form
$$\left(\matrix{\alpha_1\cr
                \alpha_2\cr
                \alpha_3\cr}\right)\propto \left(\matrix{1\cr
			                                 1\cr
        			                         \alpha_3/\alpha_1\cr}\right).
\eqno(3.36)$$
For these two modes the linearized equations reduce to
$$\left(\matrix{\cE_{11}+\cE_{12}-I_{11}\sigma^2 &\cE_{13}\cr
		2\cE_{13}&\cE_{33}-I_{33}\sigma^2\cr}\right)
  \left(\matrix{\alpha_1\cr
		\alpha_3\cr}\right)=0.
\eqno(3.37)$$
For $n\neq 0$, these two modes are somewhat similar to the homologous
pulsations
discussed in \S 3.2. They are also the compressible generalization of the
{\it zonal modes\/} discussed by Ch69 and Tassoul (1978).
Since $\alpha_3\neq\alpha_1$, the deformation is not homologous and
the results obtained in \S 3.2 are slightly modified.
In particular, setting $\sigma=0$ in equation~(3.37), we obtain
the critical $\Gamma$ for dynamical stability,
$$\Gamma_{crit}={4\over 3}-{2T_s\over 3|W|}
\left[1-\left(2+{A_3\over 3B_{13}-2A_3}\right){T_s\over |W|}\right]^{-1}.
\eqno(3.38)$$
This stability condition is slightly {\it more restrictive\/} than
equation~(3.16) since a more general trial function has been used here. With
some algebra it can be shown
that expression~(3.38) agrees with the result derived in LRS1 (eq.\ [6.3]).

The expressions for the eigenfrequencies are quite complicated for general $n$.
For $n=0$ the result is simpler. Subtracting the two equations in
(3.37) from each other, taking the $n\rightarrow 0$ limit, and using the
equilibrium conditions~(3.24), we obtain
$$\left(\matrix{2&1\cr
	I_{11}\sigma^2+(4a_1^2B_{11}-2a_3^2B_{13})\Sigma &
	(3a_3^2B_{13}-2a_3^2A_3)\Sigma-I_{33}\sigma^2 \cr}\right)
   \left(\matrix{\alpha_1\cr
		\alpha_3\cr}\right)=0,~~~~~(n=0).
\eqno(3.39)$$
The eigenfrequency is given by
$$\bar\sigma^2=\left[4B_{11}+{a_3^2\over a_1^2}(6B_{33}-4B_{13})\right]
\left({1\over 2}+{a_3^2\over a_1^2}\right)^{-1},~~~~~(n=0).
\eqno(3.40)$$
This is identical to the result given by Ch69 (eq.\ [5.63]).
The corresponding eigenmode is
$$\left(\matrix{\alpha_1\cr
                \alpha_2\cr
                \alpha_3\cr}\right)\propto \left(\matrix{1\cr
			                                 1\cr
        			                         -2\cr}\right),~~~~~(n=0).
\eqno(3.41)$$
For a nonrotating (spherical) star with $n=0$, the toroidal
mode~(3.31) is simply the $l=2$, $m=\pm 2$ Kelvin mode, while the
mode~(3.41) describes axisymmetric pulsations with $l=2$, $m=0$.
With $B_{ij}=4/15$ for a sphere, equations~(3.30) and (3.40) both give
$\bar\sigma^2=16/15$, an exact result for the frequency of the
Kelvin mode.

Note that the ``transverse-shear modes'' discussed by Ch69 (\S 33)
cannot be incorporated in our treatment, since they involve perturbations
of the rotation axis.

\bigskip
\centerline{\bf 4. DISSIPATIVE EFFECTS FOR A SINGLE STAR}
\medskip

We now incorporate dissipative forces into the dynamical equations
derived in \S 2.
In general, dissipation modifies the Euler-Lagrange equations according to
$${d\over dt}{\partial L\over\partial \dot q_i}=
{\partial L\over\partial q_i}+\cF_{q_i},\eqno(4.1)$$
where $\cF_{q_i}$ is the generalized force associated with the coordinate
$q_i$. The generalized dissipative forces are defined so that the dissipation
rate ${\cal W}$ (Rayleigh's dissipation function; cf.~Goldstein 1980) can
be written
$${\cal W}=\cF_{q_i}\dot q_i.\eqno(4.2)$$
Therefore, to calculate $\cF_{q_i}$, we need to evaluate ${\cal W}$
and express it in terms of $q_i$ and $\dot q_i$.

\bigskip
\centerline{\bf 4.1 Viscous Dissipation}
\medskip

The dissipation rate due to shear viscosity is given by
(cf.~Landau \& Lifshitz 1987)
$${\cal W}=-\int\!\!\sigma_{ij}u_{i,j}\, d^3x,\eqno(4.3)$$
where $u_i$ is the fluid velocity and $\sigma_{ij}$
is the viscous stress tensor,
$$\sigma_{ij}=\eta \left(u_{i,j}+u_{j,i}-{2\over 3}\delta_{ij}\nabla\cdot
{\bf u}\right),\eqno(4.4)$$
We denote by $\eta=\rho\nu$ the dynamical shear viscosity, where $\nu$ is the
kinematic shear viscosity. The bulk viscosity will be
neglected. From equations (2.2)--(2.4), we have
$$\eqalign{
& u_{1,1}={\dot a_1\over a_1},~~~u_{2,2}={\dot a_2\over a_2},
{}~~~u_{3,3}={\dot a_3\over a_3}, \cr
& u_{1.2}={a_1\over a_2}\Lambda-\Omega,\cr
& u_{2,1}=-{a_2\over a_1}\Lambda+\Omega,\cr
& u_{i,j}=0,~~~{\rm otherwise}.
}\eqno(4.5)$$
Thus the viscous dissipation rate is given by
$$\eqalign{
{\cal W}= &-{4\over 3}\bar\nu M\biggl [\left({\dot a_1\over a_1}\right)^2
+\left({\dot a_2\over a_2}\right)^2+\left({\dot a_3\over a_3}\right)^2
-\left({\dot a_1\over a_1}\right)\left({\dot a_2\over a_2}\right)
-\left({\dot a_1\over a_1}\right)\left({\dot a_3\over a_3}\right)\cr
&-\left({\dot a_2\over a_2}\right)\left({\dot a_3\over a_3}\right)
\biggr ]-\bar\nu M\Lambda^2\left({a_1^2-a_2^2\over a_1a_2}\right)^2,
}\eqno(4.6)$$
where $\bar\nu$ is the mass-averaged shear viscosity
$$\bar\nu={1\over M}\int\!\nu\, dm.\eqno(4.7)$$

Since ${\cal W}$ in equation (4.6) is quadratic in $\dot q_i$, from equation
(4.2),
the dissipative forces are given by
$$\cF_{q_i}={1\over 2}{\partial {\cal W}\over\partial \dot q_i}.\eqno(4.8)$$
Thus we have
$$\eqalign{
\cF_{a_1}&=-{2\over 3}\bar\nu M \left(2{\dot a_1\over a_1}
-{\dot a_2\over a_2}-{\dot a_3\over a_3}\right){1\over a_1},\cr
\cF_{a_2}&=-{2\over 3}\bar\nu M \left(2{\dot a_2\over a_2}
-{\dot a_1\over a_1}-{\dot a_3\over a_3}\right){1\over a_2},\cr
\cF_{a_3}&=-{2\over 3}\bar\nu M \left(2{\dot a_3\over a_3}
-{\dot a_1\over a_1}-{\dot a_2\over a_2}\right){1\over a_3},\cr
\cF_{\phi}&=0,\cr
\cF_{\psi}&=-\bar\nu M\Lambda\left({a_1^2-a_2^2\over a_1a_2}\right)^2.
}\eqno(4.9)$$
The statement $\cF_{\phi}=0$ simply indicates that viscous forces
conserve angular momentum ($dJ_s/dt=\cF_{\phi}= 0$),
while they do not conserve fluid circulation since $d\cC/dt=\cF_{\psi}\ne 0.$

In the presence of viscous dissipation, the dynamical equations
(2.19)--(2.23)  become
$$\ddot a_1=\{\cdots\}-{10\over 3\kappa_n}\bar\nu\left(
{2\dot a_1\over a_1}-{\dot a_2\over a_2}-{\dot a_3\over a_3}\right)
{1\over a_1},\eqno(4.10)$$
$$\ddot a_2=\{\cdots\}-{10\over 3\kappa_n}\bar\nu\left(
{2\dot a_2\over a_2}-{\dot a_1\over a_1}-{\dot a_3\over a_3}\right)
{1\over a_2},\eqno(4.11)$$
$$\ddot a_3=\{\cdots\}-{10\over 3\kappa_n}\bar\nu\left(
{2\dot a_3\over a_3}-{\dot a_1\over a_1}-{\dot a_2\over a_2}\right)
{1\over a_3},\eqno(4.12)$$
$${d\over dt}\left(a_1\Omega-a_2\Lambda\right)=\{\cdots\}
-{5\over\kappa_n}\bar\nu {a_1^2-a_2^2\over a_1^2a_2}\Lambda,\eqno(4.13)$$
$${d\over dt}\left(-a_2\Omega+a_1\Lambda\right)=\{\cdots\}
-{5\over\kappa_n}\bar\nu {a_1^2-a_2^2\over a_1a_2^2}\Lambda,\eqno(4.14)$$
where $\{\cdots\}$ denote the terms that already
exist in equations (2.19)--(2.23) (This notation will be used throughout the
paper).
For the pressure term, expression (2.30) still applies here,
but expression~(2.34) must be modified as
$${2P_c\over\rho_c}=\left(\sum_i\!\!{1\over a_i^2}\right)^{-1}
\left[\{\cdots\}+10\bar\nu\sum_i\!\!{\dot a_i\over a_i^3}\right],~~~~~~~(n=0).
\eqno(4.15)$$
Equations (2.24)--(2.25) become
$$\eqalign{
\dot\Omega &=\left({a_2\over a_1}-{a_1\over a_2}\right)^{-1}
\left[\{\cdots\}+{10\over\kappa_n}\bar\nu{a_1^2-a_2^2\over a_1^2a_2^2}\Lambda
\right],\cr
\dot\Lambda &=\left({a_2\over a_1}-{a_1\over a_2}\right)^{-1}
\left[\{\cdots\}+{5\over\kappa_n}\bar\nu{a_1^2-a_2^2\over a_1a_2}
\left({1\over a_1^2}+{1\over a_2^2}\right)\Lambda\right].\cr
}\eqno(4.16)$$
Recall that the enegy dissipation rate is simply $\dot\cE={\cal W}$.
For a quasi-static ellipsoid, this rate reduces to equation~(6.4) in LRS4.

\bigskip
\centerline{\bf 4.2 Gravitational Radiation Reaction}
\medskip

In the weak-field, slow-motion regime of general relativity,
the emission of gravitational waves induces a back-reaction scalar potential
$\Phi_{react}$ which can be written as (Misner, Thorne \& Wheeler 1970)
$$\Phi_{react}={G\over 5c^5}\Ib5_{ij}x_ix_j,\eqno(4.17)$$
where the superscript $(5)$ indicates the fifth time derivative and
$c$ is the speed of light. Here we choose $x_i$ ($i=1,2,3$) to be
the Cartesian coordinates of a fluid element
in the instantaneous corotating frame
(the body frame, with basis vectors along the principal axes). Then
$\Ib5_{ij}$ is the fifth time derivative of the reduced quadrupole moment
tensor of the system in the inertial frame
{\it projected onto the body frame}. Expressions for $\Ib5_{ij}$ are given
in Appendix~A. The radiation reaction force per unit mass on the fluid is
$-\nabla\Phi_{react}$. The energy dissipation rate is thus given by
$${\cal W}=-\int\!{\bf u}\cdot\nabla\Phi_{react}\,dm.\eqno(4.18)$$
Using equations (2.2)--(2.4), we get
$${\cal W}=-\left({1\over 5}\kappa_nM\right){2G\over 5c^5}
\left[\Ib5_{11}\dot a_1 a_1+\Ib5_{22}\dot a_2 a_2+\Ib5_{33}\dot a_3 a_3
+\Ib5_{12}\Omega(a_1^2-a_2^2)\right],\eqno(4.19)$$
where we have used
$$\int\!\! x_ix_jdm\equiv I_{ij}={1\over 5}\kappa_n Ma_i^2\delta_{ij}.
\eqno(4.20)$$

Since ${\cal W}$ here is linear in $q_i$, the dissipative forces due to
gravitational radiation are given by (compare eq.\ [4.8])
$$\cF_{q_i}={\partial {\cal W}\over\partial \dot q_i}.\eqno(4.21)$$
{}From equations~(4.19) and~(4.21) we obtain
$$\eqalign{
\cF_{a_1}&=-\left({1\over 5}\kappa_nM\right){2G\over 5c^5}\Ib5_{11}a_1,\cr
\cF_{a_2}&=-\left({1\over 5}\kappa_nM\right){2G\over 5c^5}\Ib5_{22}a_2,\cr
\cF_{a_3}&=-\left({1\over 5}\kappa_nM\right){2G\over 5c^5}\Ib5_{33}a_3,\cr
\cF_{\phi}&=-\left({1\over 5}\kappa_nM\right){2G\over 5c^5}\Ib5_{12}
(a_1^2-a_2^2),\cr
\cF_{\psi}&=0.
}\eqno(4.22)$$
Since $\cF_{\psi}=0$, we see that {\it gravitational radiation
reaction conserves the fluid circulation}, i.e.,
$d\cC/dt=\cF_\psi=0$. But since $\cF_{\phi}\ne 0$, the total angular momentum
is not conserved in general. Indeed, gravitational waves can carry away angular
momentum as well as energy when the system is not axisymmetric.

Including the gravitational radiation reaction, the dynamical equations
(2.19)--(2.23) become
$$\eqalignno{
&\ddot a_1=\{\cdots\}-{2G\over 5c^5}\Ib5_{11}a_1,  &(4.23)\cr
&\ddot a_2=\{\cdots\}-{2G\over 5c^5}\Ib5_{22}a_2,  &(4.24)\cr
&\ddot a_3=\{\cdots\}-{2G\over 5c^5}\Ib5_{33}a_3,  &(4.25)\cr
&{d\over dt}\left(a_1\Omega-a_2\Lambda\right)=\{\cdots\}
-{2G\over 5c^5}\Ib5_{12}a_1,  &(4.26)\cr
&{d\over dt}\left(-a_2\Omega+a_1\Lambda\right)=\{\cdots\}
-{2G\over 5c^5}\Ib5_{12}a_2. &(4.27)\cr
}$$

Since $\Ib5_{11}+\Ib5_{22}+\Ib5_{33}=0$, it can be shown easily
that the expression for $2P_c/\rho_c$
is not affected by the presence of gravitational radiation reaction, i.e.,
equations~(2.30) and~(2.34) still apply. Equations (2.24)--(2.25) are
modified as
$$\eqalign{
\dot\Omega &=\left({a_2\over a_1}-{a_1\over a_2}\right)^{-1}
\left[\{\cdots\}+{2G\over 5c^5}\Ib5_{12}\left({a_1\over a_2}
+{a_2\over a_1}\right)\right],\cr
\dot\Lambda &=\left({a_2\over a_1}-{a_1\over a_2}\right)^{-1}
\left[\{\cdots\}+{4G\over 5c^5}\Ib5_{12}\right].\cr
}\eqno(4.28)$$
As before the energy loss rate is simply $\dot\cE={\cal W}$.

\bigskip
\centerline{\bf 4.3 Secular Evolution of a Compressible Ellipsoid}
\medskip

The secular evolution of incompressible ellipsoids in the presence
of dissipation has been studied previously.
Using a linear perturbation analysis,
Roberts \& Stewartson (1963) and Chandrasekhar (1970)
demonstrated that the presence of either viscosity or gravitational
radiation reaction induces
a secular instability of incompressible spheroids on the Maclaurin sequence
beyond the point where the Jacobi and Dedekind sequences branch off.
Press \& Teukolsky (1973) have studied the viscous evolution of
a secularly unstable incompressible Maclaurin spheroid,
and Miller (1974) has integrated the Riemann-Lebovitz equations for
incompressible ellipsoids including the effects of gravitational
radiation reaction.

Our dynamical equations represent a generalization to
compressible ellipsoids of the equations considered by
Press \& Teukolsky (1973) and Miller (1974). We have integrated our equations
for a variety of compressible ellipsoidal configurations,
and have found results similar to those obtained in
these previous studies for incompressible ellipsoids.

The secular evolutionary paths of the ellipsoids
can be understood clearly in terms of the energetics and conservation
principles: dissipative forces
always drive a system toward a state with lower energy, along
quasi-equilibrium paths that hold conserved quantities fixed.
Consider first the evolution of an ellipsoid under gravitational
radiation reaction. Since the radiation reaction forces conserve
the fluid circulation (cf.~eq.[4.22]), such an evolution is always
along a constant-$\cC$ sequence. In Figure~1, we show the variation
of  the total energy along various constant-$\cC$ sequences (with $n=1$)
as a function of the axis ratio $a_2/a_1$.
The curves were obtained numerically by solving the Riemann-S
equilibrium equations (LRS1, \S5).
Note that a given Maclaurin spheroid corresponds to
a unique value of $\cC$, but two different
constant-$\cC$ sequences branch off:
one sequence is {\it Jacobi-like\/}, with $|\Omega|>|\Lambda|$,
the other is {\it Dedekind-like\/}, with $|\Lambda|>|\Omega|$.
\footnote{$^4$}
{Such a characterization of the two branches
(as used by Detweiler \& Lindblom 1977) is not exact.
When the deformation is very large, e.g., $a_2/a_1\le 0.1$, we find
that both configurations have $|\Omega|>|\Lambda|$.
However, except for such highly deformed configurations,
the two branches do have $|\Omega|>|\Lambda|$ or $|\Lambda|>|\Omega|$.
More appropriate is using $|\zeta/\Omega|<2$ to define Jacobi-like
and $|\zeta/\Omega|>2$ to define Dedekind-like, where $\zeta$ is the
vorticity in the corotating frame.}
For given values of $\cC$ and $a_2/a_1$, the
Jacobi-like configuration has higher energy than the Dedekind-like
configuration. Also shown in Figure~1 is
the Dedekind sequence (of configurations with $\Lambda/\Omega=\infty$) and
the Jacobi sequence ($\Lambda/\Omega=0$),
which bifurcate from the Maclaurin sequence ($a_2/a_1=1$) at the point
where the ratio $T_s/|W|=0.1375$ (in our approximation, this value is
independent of $n$; see LRS1, \S4). Note that the Dedekind
sequence and the Jacobi sequence
are adjoints of each other, and therefore have the same energy for
given $a_2/a_1$ (see LRS1, \S5.1).

We see from Figure~1 that there exists a critical value of
$|\cC|=|\cC_{sec}|$,
equal to the absolute value of the circulation of a Maclaurin spheroid at
the bifurcation point, below which a constant-$\cC$ sequence has a
Maclaurin spheroid as its minimum-energy state,
and above which its minimum-energy
state is a Dedekind ellipsoid. Therefore, an initial configuration
with $|\cC|<|\cC_{sec}|$ will evolve to become a
Maclaurin spheroid under gravitational radiation reaction.
For an initial configuration with $|\cC|>|\cC_{sec}|$, the final
state is a Dedekind ellipsoid. In this latter case, however, if
$|\Omega|>|\Lambda|$, the system will first evolve to a Maclaurin spheroid;
only when some additional perturbation (e.g., viscosity) triggers
the secularly unstable bar mode of the Maclaurin spheroid does the system
evolve past the Maclaurin sequence, toward a final Dedekind ellipsoid.
Note that the evolutionary timescale of the
Dedekind-like phase is much longer than that of the Jacobi-like phase,
as a result of the steep power-law dependence of the energy
dissipation rate on $\Omega$ (cf.\ Lai \& Shapiro 1994).

The evolution of ellipsoids under pure viscous dissipation
can be understood similarly.
As viscosity conserves $J$ (see eq.~[4.9]) while dissipating energy,
the evolution is along a
constant-$J$ sequence. From Dedekind's theorem (Ch69, Chap.~3; LRS1, \S5.1),
we know that the energy curves of constant$-J$ sequences
are identical to those of constant-$\cC$ curves, e.g.,
a Jacobi-like constant-$J$ sequence is adjoint to
a Dedekind-like constant-$\cC$ sequence with $\cC=-J$. Therefore,
Figure~1 can be used again here, but after switching the curves for Jacobi-like
sequences and Dedekind-like sequences.
We see that the final state is either a secularly stable
Maclaurin spheroid (when $J<J_{sec}$) or a Jacobi ellipsoid (when
$J>J_{sec}$).

The difference in the evolution for different polytropic indices $n$
is illustrated in Figure 2, where we show the $E$ vs $J$ curves of
Maclaurin, Jacobi and Dedekind sequences for various $n$.
In the presence of viscosity, an ellipsoid evolves vertically downward
in this diagram, and the evolution terminates at a
corresponding Maclaurin spheroid or Jacobi
ellipsoid. We see that for small $n\lo 1.5$, a Dedekind ellipsoid evolves to a
secularly stable Maclaurin spheroid; but for large $n\go 1.5$,
a Dedekind ellipsoid first evolves toward a secularly unstable
Maclaurin spheroid, and finally to a Jacobi  ellipsoid.
This qualitative difference is easy to understand: a highly
compressible configuration can expand appreciably when deformed,
giving a larger angular momentum (this can also be seen
using the scaling relation for Riemann-S ellipsoids; cf.~eq.[3.27] in LRS1).

Similarly, Figure~2 can be used to understand the evolution driven
by gravitational
radiation reaction. In this case, the horizontal axis represents $-\cC$, and
curves for Jacobi and Dedekind sequences are switched. Thus we see that
for small $n$, a Jacobi ellipsoid evolves to a Maclaurin spheroid,
while for large $n$, its final  state is a Dedekind ellipsoid.
One of the outstanding problems of 3D hydrodynamics with gravitational
radiation is to verify this behavior using the exact hydrodynamic equations.
The result has important consequences for the fate of nonaxisymmetric, rapidly
rotating neutron stars, and for coalescing binary neutron stars (see Rasio \&
Shapiro 1994, \S4.1).

Lindblom \& Detweiler (1977) have considered the combined effects of
gravitational radiation reaction and viscosity on the stability of
incompressible Maclaurin spheroids.
Based on a linear analysis, they showed that
when operating together, the two effects tend to cancel
each other. Evolutionary tracks of general ellipsoids for various
ratios of the viscous timescale and the radiation-reaction timescale
have been obtained by Detweiler \& Lindblom (1977).
Using our dynamical equations we can now extend this study to
compressible ellipsoids (Lai \& Shapiro 1994).

\bigskip
\centerline{\bf 4.4 Secular Instability Growth Time}
\medskip

We have derived the secular instability growth time for a compressible
Maclaurin spheroid, both with viscosity and gravitational radiation reaction
(Lai \& Shapiro 1994). The viscous instability growth time
$\tau_{vis}$ is given by
$$\tau_{vis}^{-1}={5\bar\nu\over\kappa_na_1^2}
\left({\Omega-\sigma\over\sigma}\right),
\eqno(4.29)$$
while the growth time $\tau_{GW}$ of the instability driven by gravitational
radiation reaction is given by
$$\tau_{GW}^{-1}={2G\kappa_nMa_1^2\over 25c^5}
{(\Omega-\sigma)^5\over\sigma}.\eqno(4.30)
$$
Here $\sigma$ is the eigenfrequency of the toroidal mode in the absence
of dissipation, given by equation~(3.30), and $\kappa_n$ is defined by
equation~(2.7).
In the the $n=0$ limit,
these results agree with those given by Ch69 (\S37) for the viscous instability
and by Chandrasekhar (1970) for gravitational radiation reaction.

A more complete discussion of the instabilities and secular evolution
of rotating stars will be presented elsewhere (Lai \& Shapiro 1994),
together with an application to the emission of gravitational waves
during core collapse.

\bigskip
\centerline{\bf 5. DYNAMICAL EQUATIONS FOR BINARIES:}
\smallskip
\centerline{\bf ROCHE-RIEMANN SYSTEMS}
\medskip

Having studied in detail the evolution of single stars modeled as
compressible Riemann-S ellipsoids, we now turn to binary systems.
We will derive the general dynamical equations for a bound or
unbound system containing a compressible Riemann-S ellipsoid
and a {\it point mass\/} (a Roche-Riemann binary system).
For parabolic orbits, Nduka (1971) first derived
the dynamical Riemann-Lebovitz equations in the incompressible limit
\footnote{$^5$}{Misprints and errors in the original paper by Nduka
have been pointed out by Luminet \& Carter (1986; p.~224)
and by Kosovichev \& Novikov (1992).}.
In addition to providing a generalization to compressible ellipsoids,
our equations also determine the orbital dynamics self-consistently, thus
allowing for general binary orbits.
Equilibrium Roche-Riemann binaries in circular orbit have been studied
in LRS1 (\S8).

In addition to the coordinates $\{a_i,\phi,\psi\}$ used in \S 2 to
describe the structure of a single ellipsoid,
we need to introduce new variables to specify the orbital motion
and the relative orientation of the ellipsoid.
For simplicity we consider only orbits {\it in the equatorial plane\/}
of the ellipsoid (i.e.,
with the orbital angular momentum and the spin of the ellipsoid aligned
along ${\bf e}_3$). Two new coordinates are then needed to describe the orbit:
a radial coordinate $r$ which measures the separation between the center of
mass of the ellipsoid and the point mass, and an angular coordinate
$\theta$ which we take to be the true anomaly of the point mass. We also
define a {\it misalignment angle\/} $\alpha=\theta-\phi$ between the axis
$a_1$ of the ellipsoid and the line joining the centers of the two bodies
(see Figure 3). The point mass is called $M'$ and, following Ch69 and LRS,
we denote the mass ratio by $p=M/M'$.

The Lagrangian of the system is simply the sum of the
single ellipsoid contribution, $L_s$, given by equations (2.8)--(2.12),
and an orbital contribution $L_{orb}$,
$$L=L_s+L_{orb},\eqno(5.1)$$
where
$$L_{orb}={1\over 2}\mu\dot r^2+{1\over 2}\mu r^2\dot\theta^2-W_i.\eqno(5.2)$$
The first two terms in equation (5.2) give the orbital kinetic energy, and
the last term $W_i$ is the interaction energy between $M$ and $M'$,
given by
$$W_i=-GM'\int\!\!d^3x{\rho({\bf x})\over |{\bf r}-{\bf x}|},\eqno(5.3)$$
where $\rho({\bf x})$ is the density distribution within $M$.
To quadrupole order, we have
$${1\over |{\bf r}-{\bf x}|}\simeq {1\over r}
+{{\bf x}\cdot{\bf \hat r}\over r^2}
+{1\over 2r^3}[3({\bf x}\cdot{\bf \hat r})^2-{\bf x}^2],\eqno(5.4)$$
where ${\bf \hat r}$ is the unit vector connecting $M$ to $M'$.
Since $\int\!({\bf x}\cdot{\bf \hat r})dm=0$, we have
$$W_i=-{GMM'\over r}-{GM'\over 2r^3}(3I_{rr}-I_{11}-I_{22}-I_{33}),\eqno(5.5)$$
where
$$I_{rr}=\int\!\!d^3x\rho({\bf x})({\bf x}\cdot{\bf \hat r})^2
=I_{11}\cos^2\alpha+I_{22}\sin^2\alpha,\eqno(5.6)$$
and $I_{ii}$ is defined by equation (4.20).
Combining (5.5) and (5.6), we obtain
$$W_i=-{GMM'\over r}-{GM'\over 2r^3}[I_{11}(3\cos^2\alpha-1)
+I_{22}(3\sin^2\alpha-1)-I_{33}].\eqno(5.7)$$

Given the Lagrangian (eqs.~[5.1] and~[5.2]), the dynamical equations can then
be
obtained from the Euler-Lagrange equations~(2.13). Of particular interest
are the equations for $\phi$ and $\theta$. For $q_i=\phi$, we get
$${dJ_s\over dt}={\cal N}={3GM'\over 2r^3}\sin 2\alpha (I_{11}-I_{22}),
\eqno(5.8)$$
where $J_s$ is the ``spin'' angular momentum of the star, given
by equation (2.17), and $\cal N$ is the tidal torque exerted on the star.
For $q_i=\theta$, equation~(2.13) gives
$${dJ_{orb}\over dt}=-{\cal N},\eqno(5.9)$$
where $J_{orb}=\mu r^2\dot\theta$ is the orbital angular momentum. Thus
we see that the total angular momentum,
$$J=J_s+J_{orb},\eqno(5.10)$$
is conserved, as expected. Equation~(2.13) for $q_i=\psi$ indicates
that the fluid circulation given by equation~(2.18) is also conserved,
since tidal forces conserve the fluid vorticity.

The complete dynamical equations for a Roche-Riemann system in the absence
of dissipation can be written as
$$\eqalignno{
&\ddot a_1 =a_1(\Omega^2+\Lambda^2)-2a_2\Omega\Lambda
-{2\pi G\over q_n}a_1A_1\bar\rho
+\left({5k_1 \over n\kappa_n}{P_c\over\rho_c}\right){1\over a_1}
+{GM'a_1\over r^3}(3\cos^2\alpha-1),&\cr
&\hfil      &(5.11)\cr
&\ddot a_2 =a_2(\Omega^2+\Lambda^2)-2a_1\Omega\Lambda
-{2\pi G\over q_n}a_2A_2\bar\rho
+\left({5k_1 \over n\kappa_n}{P_c\over\rho_c}\right){1\over a_2}
+{GM'a_2\over r^3}(3\sin^2\alpha-1),&\cr
&\hfil      &(5.12)\cr
&\ddot a_3 =-{2\pi G\over q_n}a_3A_3\bar\rho
+\left({5k_1 \over n\kappa_n}{P_c\over\rho_c}\right){1\over a_3}
-{GM'a_3\over r^3},                                                  &(5.13)\cr
&{d\over dt}\left(a_1\Omega-a_2\Lambda\right)
=-\dot a_1\Omega+\dot a_2\Lambda+{3GM'a_1\over 2r^3}\sin 2\alpha,    &(5.14)\cr
&{d\over dt}\left(-a_2\Omega+a_1\Lambda\right)
=\dot a_2\Omega-\dot a_1\Lambda+{3GM'a_2\over 2r^3}\sin 2\alpha,     &(5.15)\cr
&\ddot r =r{\dot\theta}^2-{G(M+M')\over r^2}
-{3\kappa_n G\over 10}{(M+M')\over r^4}\left[a_1^2(3\cos^2\alpha-1)
+a_2^2(3\sin^2\alpha-1)-a_3^2\right],&\cr
&\hfil     &(5.16)\cr
&\ddot \theta =-{2\dot r\dot\theta\over r}
-{3\kappa_n G\over 10}{(M+M')\over r^5}(a_1^2-a_2^2)\sin 2\alpha,
&(5.17)\cr
}$$
where $\dot\theta=\Omega_{orb}$, $\dot\phi=\Omega$.
For numerical integrations, equations (5.14)--(5.15)
can be rewritten as
$$\eqalignno{
\dot\Omega &=\left({a_2\over a_1}-{a_1\over a_2}\right)^{-1}
\left[2\left({\Omega\over a_2}+{\Lambda\over a_1}\right)\dot a_1
-2\left({\Omega\over a_1}+{\Lambda\over a_2}\right)\dot a_2
-{3GM'\over 2r^3}\left({a_1\over a_2}
+{a_2\over a_1}\right)\sin 2\alpha\right],&\cr
&\hfil      &(5.18)\cr
\dot\Lambda &=\left({a_2\over a_1}-{a_1\over a_2}\right)^{-1}
\left[2\left({\Omega\over a_1}+{\Lambda\over a_2}\right)\dot a_1
-2\left({\Omega\over a_2}+{\Lambda\over a_1}\right)\dot a_2
-{3GM'\over r^3}\sin 2\alpha\right].     &(5.19)}$$
The pressure term in equations (5.11)--(5.13) can be handled
in the same way as for a single star. It is easy to verify
that equations~(2.30) and~(2.34) still apply.

\vfill\eject
\bigskip
\centerline{\bf 6. DYNAMICAL INSTABILITY OF ROCHE-RIEMANN BINARIES}
\medskip

It is straightforward to show that, for an equilibrium
system (i.e., when $\dot a_i=0=\ddot a_i$ and $\alpha=0$), the
dynamical equations (5.11)--(5.19) reduce to the equilibrium equations
for Roche-Riemann binaries derived and solved in LRS1 (\S8.1).
We now examine the dynamical stability of the equilibrium solutions.

To study small dynamical oscillations, one could linearize the
dynamical equations
and calculate all the  eigenfrequencies. This involves extensive algebra.
However, we can show that
the onset of dynamical instability is determined by a condition similar to
equation~(3.8), with an appropriately constructed energy function.
For this purpose, it is convenient to use $\alpha$ instead of
$\theta$ as an independent variable. Thus we define
$\{q_i\}=\{a_1,a_2,a_3,r,\alpha,\phi,\psi\}$ here.
The Lagrangian of the system as discussed in \S 5 can be written as
$$\eqalign{
L =& {1\over 10}\kappa_nM(\dot a_1^2+\dot a_2^2+\dot a_3^3)
+{1\over 2}I(\dot\phi^2+\dot\psi^2)-{2\over 5}\kappa_na_1a_2\dot\phi
\dot\psi	\cr
&+{1\over 2}\mu\dot r^2+{1\over 2}\mu r^2(\dot\phi+\dot\alpha)^2-U-W-W_i.  \cr
}\eqno(6.1)$$
The canonical momenta associated with the $\{q_i\}$ are
$$P_{a_i}={1\over 5}\kappa_nM\dot a_i,~~~
P_r=\mu \dot r,~~~P_\alpha=\mu r^2(\dot\phi+\dot\alpha),
{}~~~P_{\phi}=J,~~~P_{\psi}=\cC.\eqno(6.2)$$
Because of the conservation of $J$ and $\cC$, and also because
$\alpha=0$ for an equilibrium configuration, it is convenient to define a
subset of the variables $\{\alpha_i\}\equiv\{a_1,a_2,a_3,r\}$. The Hamiltonian
can then be written as
$$H(\alpha_i,\alpha,P_{\alpha_i},P_\alpha,J,\cC)
= {1\over 2(\kappa_nM/5)}(P_{a_1}^2+P_{a_2}^2+P_{a_3}^2)
+{1\over 2\mu}P_r^2+E(\alpha_i,\alpha,P_\alpha,J,\cC),
\eqno(6.3)$$
where
$$E(\alpha_i,\alpha,P_\alpha,J,\cC)={1\over 2\mu r^2}P_\alpha^2+
{1\over 2I_{+}}(J-P_\alpha+\cC)^2+{1\over 2I_{-}}(J-P_\alpha-\cC)^2+U+W+W_i,
\eqno(6.4)$$
(compare with eq.~[3.3]), where $I_{\pm}$ is defined in equation (3.4).
Hamilton's equations $dP_{q_i}/dt=-\partial H/\partial q_i$ for $q_i=a_i$
give
$${1\over 5}\kappa_nM\ddot a_i
=-\left({\partial E\over \partial a_i}\right)_{P_\alpha,J,C}
=-\left({\partial E\over \partial a_i}\right)_{J,C}
+\dot\alpha\left({\partial P_\alpha\over\partial a_i}\right)_{J,C},
\eqno(6.5)$$
where for the second equality we have used $\partial E/\partial P_\alpha
=\dot\alpha$, and we consider $P_\alpha$ to be
a function of $\alpha_i,~J,~\cC$ and $\dot\alpha$, and
$E$ to be a function $E(\alpha_i,\alpha,\dot\alpha,J,\cC)$.
Linearization of equation~(6.5) about equilibrium, letting
$\alpha_i=\alpha_{i,eq}+\delta \alpha_i$, yields
$$\eqalign{
{1\over 5}\kappa_nM\delta \ddot a_i &=
-\sum_j\delta\alpha_j\left({\partial^2E\over\partial a_i\partial\alpha_j}
\right)_{eq}
-\alpha\left({\partial^2E\over\partial a_i\partial\alpha}\right)_{J,C}
-\dot\alpha\left({\partial^2E\over\partial a_i\partial\dot\alpha}\right)_{J,C}
+\dot\alpha\left({\partial P_\alpha\over\partial a_i}\right)_{J,C}	\cr
&=
-\sum_j\delta\alpha_j\left({\partial^2E\over\partial a_i\partial\alpha_j}
\right)_{eq}
-\dot\alpha\left({\partial^2E\over\partial a_i\partial\dot\alpha}\right)_{J,C}
+\dot\alpha\left({\partial P_\alpha\over\partial a_i}\right)_{J,C},	\cr
}\eqno(6.6)$$
where the second equality follows because $(\partial^2E/\partial
a_i\partial\alpha)\propto \sin\alpha$ (see eq.~[5.8]) and $|\alpha|\ll1$ near
equilibrium. Similarly, Hamilton's equation for $q_i=r$ yields
$$\mu\delta\ddot r=
-\sum_j\delta\alpha_j\left({\partial^2E\over\partial r\partial \alpha_j}
\right)_{eq}
-\dot\alpha\left({\partial^2E\over\partial r\partial\dot\alpha}\right)_{J,C}
+\dot\alpha\left({\partial P_\alpha\over\partial r}\right)_{J,C}.
\eqno(6.7)$$
Now we substitute $\delta\alpha_i\propto e^{i\sigma t}$ in
equations~(6.6) and~(6.7) and let $\sigma=0$. We obtain
$$\sum_j \delta\alpha_j\left({\partial^2E\over\partial
\alpha_i\partial\alpha_j}\right)_{eq}=0,~~~~~{\rm for}~~\sigma=0.
\eqno(6.8)$$
Clearly, in this expression, $E$ should be evaluated at $\alpha=0$, the
equilibrium value. Thus we
can write $E=E(\alpha_i; J,\cC)$ with fixed $\alpha=0$, and
the onset of dynamical instability is determined from the condition
$${\rm det}\biggl({\partial^2E\over\partial \alpha_i\partial \alpha_j}
\biggr)_{eq}=0,~~~~i,j=1,2,\ldots~~~~({\rm onset~of~instability})
	\eqno(6.9)$$
where the partial derivatives are evaluated holding $J,~\cC$
fixed. This condition forms the basis of the stability analysis of binary
systems presented in LRS1 and LRS4. Although the new degree of freedom
associated with $\alpha$ was not introduced in our previous analyses, the
explicit derivation given above shows that the stability condition
is not affected.

Alternatively, on quite general grounds,
one can show that the stability conditions determined from
equation (6.9) coincide with {\it turning points\/} along
appropriately constructed equilibrium sequences (LRS1, \S2.3).
Specifically, a dynamical stability limit coincides with the point where
the total equilibrium energy and angular momentum are both minimum
along a sequence with constant circulation.
Using both methods (eq.~[6.9] and the turning point method),
we have found in LRS1 (\S9.2) that a Roche-Riemann binary can become
dynamically unstable when the orbital separation is sufficiently small.
This instability results from the strong tidal interaction, which can
make the effective interaction potential between the two stars
much steeper than $1/r$, thereby destabilizing a circular orbit
(cf.\ Goldstein 1980, \S 3-6; see also LRS2 for a qualitative discussion).
The stability limits for various Roche-Riemann binary models
have been tabulated in LRS1 (Tables~10 and~11).

To illustrate how the instability develops, using our dynamical equations,
we show in Figure~4 the time evolution of an unstable system with
$n=1$, $p=M/M'=1$ and $\Lambda=0$ (corotating).
The dynamical equations were integrated numerically using a standard
fifth-order Runge-Kutta scheme with adaptive stepsize (Press et al 1992).
At $t=0$, an equilibrium solution is constructed for $r/a_1=1.7$,
and $r/R_o=2.406$. This equilibrium solution is then perturbed by
setting $\dot r=10^{-3}(GM/R_o)^{1/2}$. For comparison, the results
of an integration for a stable binary with $r/a_1=1.8$, $r/R_o=2.427$, and
with the same applied perturbation is also shown.
The dynamical stability limit along the corotating (Roche) sequence with
$n=1$ and $p=1$ is at $r/a_1=1.760$, or $r/R_o=2.417$.
We see clearly in Figure~4 that, as the dynamical instability
develops, $a_1$ increases while $r$ decreases, and this is accompanied
by the significant development of a tidal lag $\alpha>0$.
Of course, the precise evolution of an unstable binary depends on
how the initial configuration is perturbed.

\bigskip
\centerline{\bf 7. TIDAL CAPTURE AND DISRUPTION OF A STAR}
\smallskip
\centerline{\bf BY A MASSIVE BODY}
\medskip

Tidal interactions of stars and other fluid bodies have been discussed
extensively in a number of different contexts.
Fabian, Pringle and Rees (1975) originally proposed that tidal
encounters between a neutron star and a main-sequence star
might lead to the formation of X-ray binaries in globular clusters
(but see Rasio \& Shapiro 1991; Kochanek 1992a; Rasio 1993).
The first quantitative calculations of the orbital energy dissipation
were performed by Press \& Teukolsky (1977, hereafter PT) using a
linear perturbation method.
They were followed by many other studies using both linear theory
(Lee \& Ostriker 1986; McMillan, McDermott \& Taam 1987;
Kochanek 1992a and references therein) and numerical hydrodynamic
calculations (Rasio \& Shapiro 1991). The tidal disruption
of stars by a massive black hole can provide a mechanism for
fueling low-luminosity
active galactic nuclei (AGN), and may also lead to observable flares
in the luminosity of AGN. Many aspects of this problem
have been considered previously (Hills 1975; Rees 1988; Evans \&
Kochanek 1989; Carter \& Luminet 1985; Novikov, Pethick \& Polnarev 1992),
including relativistic effects (Laguna et al.\ 1993). Tidal
disruption of small
bodies (planetesimals) by protoplanets has also been discussed
in the context of the Solar System formation
(Boss, Cameron \& Benz 1991; Sridhar \& Tremaine 1992).

The advantage of our ellipsoidal method for studying this
problem is that it allows for the treatment of nonlinear effects, which are
inevitable for close encounters. Using their affine stellar model,
Carter and Luminet (1985, 1988) have studied in detail
the tidal disruption of stars by massive black holes.
Other studies using similar affine-type models include
those of Kochanek (1992a), Kosovichev \& Novikov (1992), and
Sridhar \& Tremaine (1992). There are several inconsistencies
and differences in the results obtained in those previous studies.
Thus we consider it worthwhile to re-examine the problem using
our independent formulation of the dynamics, even though
our equations are formally equivalent to those of the affine model.
We include in some of our calculations the effects of fluid viscosity,
which can be significant for the viscoelastic material in planetesimals
(Sridhar \& Tremaine 1992).

For definiteness, we focus on the encounter of a
fluid body of mass $M$ with a point-like object of mass
$M'\gg M$, and we consider only parabolic trajectories.
A useful dimensionless parameter characterizing the encounter is
$$\eta=\left({M\over M+M'}\right)^{1/2}\left({r_p\over R_o}\right)^{3/2},
\eqno(7.1)$$
where $r_p$ is the periastron separation.
The quantity $\eta$ is simply the ratio of the
timescale for the periastron passage, $\sim r_p/v_p$,
where $v_p$ is the velocity at
the periastron, to the dynamical timescale of the star,
$\sim R_o^{3/2}/(GM)^{1/2}$.
When the relative velocity $v_\infty$ between $M$ and $M'$ at infinite
separation is nonzero, the orbit is slightly hyperbolic. However, as long as
$v_p\gg v_\infty$, the trajectory remains very nearly parabolic close to
periastron, where the tidal interaction is strongest.

\vfill\eject
\bigskip
\centerline{\bf 7.1 Dynamical Calculations for Compressible Ellipsoids}
\medskip

For a given $\eta$, we integrate equations (5.11)--(5.19) numerically
to determine the dynamical effects of the tidal interaction.
Initially the star is placed on a parabolic orbit, far away from the
massive body ($r/R_o\gg\eta^{2/3}p^{-1/3}$),
and it is assumed to be spherical (nonrotating).
To avoid the singularity in equations (5.18)--(5.19) when
$a_1=a_2$, the initial configuration is slightly perturbed
by increasing (decreasing) $a_1$ ($a_2$) by $\sim0.1\%$. We have checked
that the final numerical results (including the energy transfer)
are independent of the exact amplitude of this initial perturbation.
The total angular momentum, fluid circulation, and energy are conserved
to very high accuracy throughout the evolution (typical error $\lo10^{-7}$).
For $n=0$, we also check the conservation of the volume: the product
$a_1a_2a_3$ remains constant to within $\sim10^{-7}$ typically.
For most calculations we use $M'/M=10^6$, but the results are
essentially independent of the exact value as long as $M'/M\go 10^3$.

Typical results are illustrated in Figure~5 for an encounter
with $\eta=2.8$ and $n=1.5$. Here $t=0$ corresponds to the time of
periastron passage. After the encounter, the star becomes an oscillating
Riemann-S ellipsoid with zero fluid circulation, but
finite angular momentum. The angular momentum deposited in the
star through the tidal interaction does not lead to
uniform spin in the absence of fluid viscosity.

\bigskip
\centerline{\bf 7.2 Energy and Angular Momentum Transfer in the Linear Theory}
\medskip

The energy transferred from the orbit to the star during a tidal encounter
can be calculated exactly in the limit of linear perturbations
using the theory developed by Press \& Teukolsky (1977). Here we also
calculate the corresponding transfer of angular momentum using the
same formalism.

The amount of energy transferred can be written as
$$\Delta E=-\int\!dt\int\!d^3x\,\rho\, {\partial \bxi\over\partial t}
\cdot\nabla {\cal U}, \eqno(7.2)$$
where ${\cal U}$ is the gravitational potential of the point mass,
${\cal U}({\bf x},t)=-{GM'/|{\bf x}-{\bf r}|}$, with
${\bf r}$ describing the orbital trajectory and ${\bf x}$ specifying
the position of a fluid element inside $M$.
The Lagrangian displacement $\bxi$ of a fluid element in the star
is assumed to be small, and can be decomposed into normal
eigenmode components $\bxi_{nlm}$, where $\{n,l,m\}$
are indices specifying the eigenmodes.
An important quantity in any discussion of tidal interactions is the
dimensionless coefficient characterizing the coupling between the tidal
potential and a particular normal mode (Zahn 1970; PT),
$$Q_{nl}=\int\!\!d^3x\,\rho\,\bxi_{nlm}^\ast\cdot\nabla
\left[r^lY_{lm}(\theta,\phi)\right]
=\int_0^R\!\rho l r^{l+1}dr\left[\xir_{nl}(r)+(l+1)\xip_{nl}(r)\right],
\eqno(7.3)$$
where, for spheroidal modes, we have expressed $\bxi_{nlm}$ as a sum of
radial and tangential components,
$$\bxi_{nlm}(\br)
=[\xir_{nl}(r)\be_r+r\xip_{nl}(r)\nabla]Y_{lm}(\theta,\phi).
\eqno(7.4)$$
The normal modes $\bxi_{nlm}$ are normalized so that
$\int\!d^3x\rho\bxi_{nlm}\cdot\bxi_{n'l'm'}=\delta_{nn'}
\delta_{ll'}\delta_{mm'}$.
The total energy transfer during a parabolic encounter can be written (PT)
$$\Delta E={GM'^2\over R_o}\sum_{l=2}^\infty\left({R_o\over r_p}\right)^{2l+2}
T_l(\eta),\eqno(7.5)$$
where the dimensionless function $T_l$ is given by
$$T_l(\eta)=2\pi^2\sum_n|Q_{nl}|^2\sum_{m=-l}^l|K_{nlm}|^2.\eqno(7.6)$$
The function $K_{nlm}$ involves the eigenfrequencies $\omega_{nl}$ and
is given in PT.

The angular momentum transfer during an encounter is given by
$$\Delta J_z=\int\!dt\int\!d^3x\,(\rho+\delta\rho) \tau_z
=\int\!dt\int\!d^3x\,\delta\rho\left(-{\partial
{\cal U}\over\partial\phi}\right),
\eqno(7.7)$$
where $\delta\rho=-\nabla\cdot (\rho\bxi)$ is the Eulerian perturbation of
the fluid density in the star, and $\tau_z=-\partial {\cal U}/\partial\phi$ is
the tidal torque per unit mass. Using a similar procedure as in PT,
we obtain (Lai 1994b)
$$\Delta J_z={GM'^2\over R_o}\left({GM\over R_o^3}\right)^{-1/2}
\sum_{l=2}^\infty\left({R_o\over r_p}\right)^{2l+2} S_l(\eta),\eqno(7.8)$$
where the dimensionless function $S_l$ is given by
$$S_l(\eta)=2\pi^2\sum_n{|Q_{nl}|^2\over\omega_{nl}}\sum_{m=-l}^l
(-m)|K_{nlm}|^2.\eqno(7.9)$$
Note that $K_{nlm}$ is larger for
$m=-2$, thus $\Delta J_z$ in (7.8) is positive.
Also note that contributions to $\Delta E$ and to $\Delta J$ from different
terms in the sum are related by
$\Delta J_{nlm}=(-m/\omega_{nl})\Delta E_{nlm}.$
\footnote{$~^6$}
{In eqs.~(7.6) and (7.9), the index $m$ does not correspond to the original
mode index in $Y_{lm}$. Instead, contributions
from $m=2$ and $m=-2$ modes have been re-grouped to derive these final
expressions. In fact, it can be shown that the $m=2$ and $m=-2$ modes
contribute to the tidal energy and angular momentum equally.}

In the incompressible limit ($n=0$ and $\Gamma=\Gamma_1
=\infty$), only f-modes of oscillation exist. These have eigenfrequencies
$\omega_{0l}^2/(GM/R^3)=2l(l-1)/(2l+1)$ (see, e.g., Cox 1980).
For the dominant $l=2$ quadrupole modes (including the Kelvin mode, cf.\
\S3.3),
$\omega_{02}^2/(GM/R^3)=4/5$. The normalized eigenfunctions are
$\xir_{02}=2\xip_{02}=(8\pi/3)^{1/2}r$, for which we have
$Q_{02}=(3/2\pi)^{1/2}MR^2$.

\vfill\eject
\bigskip
\centerline{\bf 7.3 Large-$\eta$ Encounters: Comparison with Linear Theory}
\medskip

In Figure~6, we compare our ellipsoidal results for the energy and angular
momentum transfer (\S7.1) with those of linear theory.
In the limit where $\eta\gg1$, the linear theory
should be exact. The three values of $n$ we have considered are $n=0$, 1.5,
and 2.5. For $n=0$, one would expect the results from the two methods
to agree precisely, since only the f-modes are involved in the linear theory,
and the $l=2$ quadrupole interaction is the leading order
(the $l=2$ mode structure obtained from the two methods is the
same; see \S 3.3). We find that this is not the case,
i.e., even in the limit of $\eta\rightarrow
\infty$, the results from the two methods do not agree.
The same conclusion was reached by
Kosovichev \& Novikov (1992). Kochanek (1992a) considered only
$n=1.5$ and concluded that the discrepancy was probably due to small
errors in the numerical integration of his affine-model equations.
Kosovichev \& Novikov (1992) attributed the discrepancy to slight
deviations of the true f-mode displacement from an exact ellipsoid.
They argued that in an ellipsoidal model, the stellar shape is too prolate,
leading to stronger tidal interaction. We question the validity of this
interpretation, since for a circular binary in equilibrium, it can be
shown explicitly that the energy shifts obtained from the linear theory
and from the ellipsoidal (Roche-Riemann) model are identical (Lai 1994b).
Note, however, from Figure~6, that the difference in the amount of energy
dissipated is less than $40\%$. The resulting tidal capture radius is therefore
hardly affected, since the dependence of $\Delta E$ on $r_p$ is extremely
steep.

In the compressible cases ($n> 0$), other types of nonradial
oscillation modes, which are absent in the ellipsoidal model,
can be excited by the tidal interaction. As discussed in \S 2.1, the
ellipsoidal model incorporates only the $l=2$ f-modes of oscillation.
In general, p-modes are also excited, and, if $\Gamma_1>\Gamma$,
one should also
include g-modes. However, the coupling coefficients $Q_{nl}$ (eq.~[7.3])
of the p-modes and g-modes are typically much smaller than that of the
f-modes (Lee \& Ostriker 1986).
When g-modes exist ($\Gamma_1>\Gamma$),
their contribution to the tidal energy transfer is larger than
that of the f-modes at large separation
(see the dotted lines in Fig.~6). The reason is that at larger separation,
``resonances'' occur since the (very low) frequency of the tidal driving force
is always close to that of some high-order g-mode. These resonances
lead to a more efficient energy transfer at large separation.

\bigskip
\centerline{\bf 7.4 Close Encounters: The Tidal Disruption Limit}
\medskip

As $\eta$ decreases, the amplitude of the tidal perturbations
increases and linear theory eventually breaks down. Below
the tidal {\it disruption limit\/} at some critical $\eta=\eta_{dis}$,
the amount of energy transferred to the star exceeds its original binding
energy (the stellar energy $E_s$, eq.~[2.26], becomes positive),
and the star is left unbound after the encounter.
In Figure~7 we show the results of two calculations for $n=1.5$.
For one encounter we used
$\eta=1.85$, slightly larger than the disruption limit $\eta_{dis}=1.84$,
and for the other we used $\eta=1.83<\eta_{dis}$.
For $\eta>\eta_{dis}$, the star remains bound after
the periastron passage, although the axes experience large-amplitude nonlinear
oscillations when $\eta$ is very close to $\eta_{dis}$.
For $\eta<\eta_{dis}$, the star becomes unbound ($E_s>0$); at least one of
the axes keeps increasing monotonically in time after the encounter,
leading to a progressively more and more elongated structure.
This behavior is identical to that found by Kosovichev \& Novikov (1992)
for $n=0$.

In Table~2, we list the values of $\eta_{dis}$ for different polytropic
indices. For $n=0$, our results agree precisely with those of
Kosovichev \& Novikov (1992) and Sridhar \& Tremaine (1992),
but not with those of Luminet \& Carter (1986).
For larger $n$, $r_{dis}$ decreases, since the tidal interaction is
less effective for more centrally concentrated
objects. For comparison, we also list in Table~2
the absolute {\it Roche-Riemann limit\/}, $\eta_{rr}$,
which corresponds to the minimum separation for an equilibrium configuration,
among all Roche-Riemann binaries,
to exist in a circular orbit (see LRS1, \S 8.2). Note that the Roche limit
for corotating binaries, as well as the irrotational Roche-Riemann limit
for $\cC=0$ binaries (LRS3), all correspond to larger separations
than $\eta_{rr}$.
{}From Table~2 we see that for all cases, $\eta_{rr}>\eta_{dis}$, which is
intuitively expected.

We note that even when $\eta>\eta_{dis}$, the object can become highly
elongated
(see Figure~7), with $a_1\gg a_2, a_3$, even though on simple energetic
grounds it is still bound. In
reality, such needle-like object is likely to be subject to the
``sausage'' instability of infinite cylinders (Chandrasekhar 1961),
and would break up into small pieces. Of course,
our ellipsoidal model is not capable of treating this process.

\bigskip
\centerline{\bf 7.5 Effects of Viscosity}
\medskip

Viscous dissipation forces can be easily incorporated into the
dynamical equations (5.11)--(5.19).
Since the motion of the center of mass of the star is not affected by
viscous dissipation (which depends only on the shear stresses inside the star),
the viscous forces derived for an isolated star (\S 4.1) can be directly
applied
to binaries. The dynamical equations (5.11)--(5.15) are modified in exactly
the same way as in equations (4.10)--(4.16) for
single stars. Also, since $\cF_r=\cF_\theta=0$, equations
(5.16)--(5.17) remain unchanged.
We assume that ${\bar\nu}$ is a constant during the dynamical evolution.

Typical results are illustrated in Figure~8. Here the fluid viscosity
has the value $\bar\nu=0.01\,(GMR_o)^{1/2}$.
We find that the disruption limit $\eta_{dis}$ is smaller than
in the inviscid case.
For $\eta>\eta_{dis}\simeq 1.72$, the star is still bound after the
encounter ($E_s<0$). Comparing with Figure~7, we see that the fluid viscosity
damps out the large-amplitude, nonlinear oscillations of the ellipsoid after
the encounter. Since the fluid circulation is not conserved
in the presence of viscosity, the fluid does {\it not\/} remain irrotational.
Instead, because viscous forces tend
to damp out the differential rotation (i.e., $\Lambda\rightarrow 0$ after
the encounter), the star evolves to become an equilibrium Jacobi ellipsoid
on the viscous dissipation timescale. This is very different from the
undamped, inviscid case. When the post-encounter stellar energy
$E_s$ is positive, as in the case where $\eta=1.71$ in Figure~8,
the star becomes unbound, with one of the axes
increasing monotonically. This is similar to what we found in the
inviscid case.

In Figure~9, we show the disruption limits as a function of the fluid
viscosity for three values of $n$. The maximum physical value for the
viscosity is $\bar\nu\approx(GMR_o)^{1/2}$,
corresponding to momentum transport
across the whole star in a dynamical timescale. In all cases,
the disruption limit
of a viscous body is smaller than that of inviscid body, as a result of
the viscous damping of kinetic energy. This result is in
agreement with the conclusions reached by Sridhar \& Tremaine (1992).
Note, however, that the reduction of $\eta_{dis}$
is significant only for very large viscosities,
$\bar\nu\go10^{-2}\,(GMR_o)^{1/2}$.
Typical viscosities in stars (even the large turbulent viscosity in convective
envelopes) always remain $\lo 10^{-4}\,(GMR_o)^{1/2}$.

\bigskip
\centerline{\bf 8. SECULAR EVOLUTION OF BINARIES DRIVEN BY VISCOSITY}
\medskip

As discussed in \S 7.5, the viscous dissipation forces can be easily
incorporated into the dynamical equations for the binaries. In this
section, we consider the secular evolution driven by viscous dissipation
of a binary in a circular orbit.
This subject has been well studied in the literature (see, e.g., Goldreich
\& Peale 1968 for a review). The basic ingredients were already
established in the the weak friction theory developed by G.~Darwin
more than a century ago.
Our purpose in this section is to consider some aspects
of this classical astronomical problem in the context of our compressible
ellipsoid model, which naturally extends the early treatments
 to the nonlinear regime.

\bigskip
\centerline{\bf 8.1 Tidal Lag Angle Due to Viscosity}
\medskip

Our present understanding of the tidal evolution of binary systems
is largely based on the weak friction model. In this model,
the viscosity of the star induces a lag angle between the static
tidal bulge and the direction to the companion. This results in a
torque and angular momentum transfer between the spin of the star
and the orbit. This mechanism can lead to synchronization of the
spin with the orbital motion, and a corresponding evolution of the binary
orbit. Any degree of nonsynchronization is necessarily associated
with a tidal lag angle, given by
$$\alpha\sim {\Delta\Omega\over \omega_o^2t_{visc}}
\sim {\bar\nu R\Delta\Omega\over G M},\eqno(8.1)$$
where $\Delta\Omega=\Omega-\Omega_s$, $\omega_o\sim (GM/R^3)^{1/2}$
is the fundamental frequency of the star, and $t_{visc}\sim R^2/\nu$
is the viscous dissipation time.

We can easily derive the exact result in the limit of large binary
separation using our ellipsoidal model.
For large $r/R$ we have $a_1\simeq a_2$
so that $J_s\simeq -\cC$ (see eqs.~[2.17]-[2.18]).
Using equation (5.8) for $dJ_s/dt$ and $d\cC/dt=-{\bar\nu}M\Lambda
(a_1^2-a_2^2)^2/(a_1a_2)^2$ (see eq.~[4.9]), we find
$$\sin 2\alpha\simeq
\bar\nu\Lambda {10r^3\over 3\kappa_n G M'R^2}\left({a_1^2-a_2^2\over
a_1a_2}\right).\eqno(8.2)$$
For large $r$ we have
$${a_1^2-a_2^2\over a_1a_2}\simeq {15\over 2}q_n{M'\over M}
\left({R\over r}\right)^3,\eqno(8.3)$$
(see eq.~[A25] in LRS4), and the tidal lag becomes
$$\alpha \simeq {10\over 4}(5-n){\bar\nu \Lambda R\over G M},\eqno(8.4)$$
providing the coefficient of proportionality in equation~(8.1)
(recall that $\Lambda\simeq\Omega-\Omega_s$).

\bigskip
\centerline{\bf 8.2 Viscous Evolution}
\medskip

In the presence of viscosity, only synchronized binaries can be in true
equilibrium. Any degree of nonsynchronization necessarily leads to
evolution. However, when the viscosity is sufficiently small,
the binaries evolve slowly along a sequence of quasi-equilibrium
configurations. Thus, in most cases,
dynamical calculations are not needed to follow the viscous evolution.
Since the viscous forces conserve angular momentum,
the binary evolution driven
by viscosity proceeds along sequences of constant total
angular momentum $J$. This has been discussed extensively in LRS4 (\S6).
Depending on the value of $J$, the final fate of the binary can be
qualitatively
different. Consider the example in Figure~10, where we show the
variation of the total energy along three different equilibrium sequences:
one is the corotating sequence, and
the other two have constant $J$.
All three sequences are for $n=0$ and $M/M'=1$.
A critical angular momentum
$J=J_{crit}$ is the minimum angular momentum along the corotating sequence.
When $J>J_{crit}$, the constant-$J$ curve intersects the corotating curve,
at two points corresponding to the minimum or maximum of the
energy along of the constant-$J$ sequence. For $J<J_{crit}$,
no configuration along the constant-$J$ sequence lies on the
corotating sequence.
Clearly, for $J>J_{crit}$, the viscous forces drive the binary toward a stable
corotating configuration, while for $J<J_{crit}$, the viscous forces
drive orbital decay to final coalescence.
More detailed discussions of these points can be found in LRS4.
Values of $J_{crit}$ for general $n$ and $p$ can be found in Table~10 of LRS1.

When a dynamical instability is encountered during secular orbital
decay, dynamical calculations are needed to follow the subsequent
evolution. In Figure~11 we show an example of such a dynamical calculation.
Here $p=M/M'=1$, $n=1$, and the viscosity $\bar\nu=0.01\,(GMR_o)^{1/2}$
is assumed to be constant throughout the evolution.
The initial state is constructed by solving the equations for
an equilibrium Roche-Riemann configuration with
$f_R\equiv -\Lambda (a_1^2+a_2^2)/(\Omega a_1a_2)=-4$ and $r/a_1=3$
(LRS1, \S8), corresponding to a total angular momentum $J=1.220$.
We set $\alpha=0$ at $t=0$.
After a transient oscillation, $\alpha$ attains a small
but finite value, corresponding to the tidal lag derived in \S 8.1.
The reason is that, as mentioned before, in the presence of viscosity,
the binary is not exactly in equilibrium state, and and viscosity-induced
tidal lag is inevitable. During the secular evolution, the binary stays very
close to an equilibrium evolution track with constant $J$.
However, as the dynamical stability limit
is approached, the evolution becomes much faster, and a large
{\it dynamical\/} tidal lag develops. This dynamical behavior
is similar to that shown in Figure~4.

\bigskip
\centerline{\bf 9. BINARY EVOLUTION DRIVEN BY GRAVITATIONAL}
\smallskip
\centerline{\bf RADIATION REACTION}
\medskip

In this section, we incorporate gravitational radiation reaction in our
dynamical equations for binaries. This allows us to study
binary coalescence driven by the emission of gravitational waves.
In \S 9.2 we consider the coalescence of a Roche-Riemann binary.
In particular, we  study how the dynamical instability is approached
as the binary orbit decays.
Applications to coalescing neutron star binaries will be presented
in a forthcoming paper (Lai, Rasio, \& Shapiro 1994c).

\bigskip
\centerline{\bf 9.1 Dynamical Equations Including
Gravitational Radiation Reaction}
\medskip

To calculate the gravitational radiation reaction forces, we consider
two coordinate systems (see Fig.~3): the
{\it body coordinate system\/}, centered on $M$, with basis vectors
$\{\bfe_i\}$
spanning the principal axes, and the {\it orbital coordinate system\/},
centered at the CM of the system, with basis vectors $\{\bfe_\bi\}$;
$\bfe_{\bone}$ is along the line joining $M$ and $M'$, $\bfe_\btwo$
is perpendicular to $\bfe_\bone$ in the orbital plane, and $\bfe_\bthr$
is perpendicular to the orbital plane. The two coordinate systems are related
by
$$\eqalign{
x_\bone &=x_1\cos\alpha+x_2\sin\alpha-\rcm,\cr
x_\btwo &=-x_1\sin\alpha+x_2\cos\alpha,\cr
x_\bthr &=x_3.
}\eqno(9.1)$$

We now write the gravitational radiation back-reaction potential as
$$\Phi_{react}={G\over 5c^5}\Ib5_{\bi\bj}x_\bi x_\bj.\eqno(9.2)$$
This has the same from as equation (4.17), except that here we have used
a different coordinate system: $\Ib5_{\bi\bj}$ is the fifth derivative of
the reduced quadrupole moment
tensor of the system {\it projected onto the orbital frame\/}.
Expressions for $\Ib5_{\bi\bj}$ are derived in Appendix A.

For $M'$, the velocity is simply
$${\bf v}'={\dot r}_{cm}'\bfe_\bone+\Omega_{orb}r_{cm}'\bfe_\btwo,\eqno(9.3)$$
where $r_{cm}'$ is the distance between CM and $M'$.
Thus the contribution to the dissipation function from $M'$ is
$${\cal W}_{M'}=-{2G\over 5c^5}
[M'r_{cm}'{\dot r}_{cm}'\Ib5_{\bone\bone}
+M'r_{cm}'^2\Omega_{orb}\Ib5_{\bone\btwo}].
\eqno(9.4)$$

The fluid velocity in $M$ can be written as
$${\bf v}={\bf u}+{\bf u}_{orb}=u_i\bfe_i
+(-{\dot r}_{cm}\bfe_\bone-\Omega_{orb}\rcm\bfe_\btwo),\eqno(9.5)$$
where ${\bf u}$ is the fluid velocity relative to the CM of $M$ and
${\bf u}_{orb}$ is the orbital velocity.
{}From equations (2.2)--(2.4), we have
$$\eqalign{
u_1&={\dot a_1\over a_1}x_1+\left({a_1\over a_2}\Lambda-\Omega\right)x_2,\cr
u_2&={\dot a_2\over a_2}x_2+\left(-{a_2\over a_1}\Lambda+\Omega\right)x_1,\cr
u_3&={\dot a_3\over a_3}x_3.
}\eqno(9.6)$$
Therefore,
$${\bf v}\cdot\nabla\Phi_{react}=
[u_k\bfe_k+{\bf u}_{orb}]\cdot{2G\over 5c^5}\Ib5_{\bi\bj}x_\bj\bfe_\bi.
\eqno(9.7)$$
Using equation (9.5) and the relation between $\{\bfe_i\}$ and $\{\bfe_\bi\}$,
we get
$$\eqalign{
{\bf v}\cdot\nabla\Phi_{react}={2G\over 5c^5}\biggl [
& u_1x_\bj\left(\Ib5_{\bone\bj}\cos\alpha-\Ib5_{\btwo\bj}\sin\alpha\right)
+u_2x_\bj\left(\Ib5_{\bone\bj}\sin\alpha+\Ib5_{\btwo\bj}\cos\alpha\right)\cr
&+u_3x_\bj\Ib5_{\bthr\bj}
-{\dot r}_{cm}x_\bj\Ib5_{\bone\bj}-\rcm\Omega_{orb}x_\bj\Ib5_{\btwo\bj}
\biggr].\cr
}\eqno(9.8)$$
Substituting equations (9.1) and (9.6) in this expression,
integrating over the mass distribution in $M$, and adding the
contribution from $M'$ (eq.~[9.4]), we obtain the total dissipation rate
$$\eqalign{
{\cal W} = & {\cal W}_M+{\cal W}_{M'}
   =-\int_M\!{\bf v}\cdot\nabla\Phi_{react}\, dm+W_{M'}\cr
= &-{2G\over 5c^5}\left({1\over 5}\kappa_nM\right )\biggl [
\left(\Ib5_{\bone\bone}\cos^2\alpha+\Ib5_{\btwo\btwo}\sin^2\alpha
-\Ib5_{\bone\btwo}\sin 2\alpha\right)a_1\dot a_1 \cr
&+\left(\Ib5_{\bone\bone}\sin^2\alpha+\Ib5_{\btwo\btwo}\cos^2\alpha
+\Ib5_{\bone\btwo}\sin 2\alpha\right)a_2\dot a_2
+\Ib5_{\bthr\bthr}a_3\dot a_3 \cr
&+\left(\Ib5_{\bone\btwo}\cos 2\alpha
+(\Ib5_{\bone\bone}-\Ib5_{\btwo\btwo}){1\over 2}\sin 2\alpha\right)
(a_1^2-a_2^2)\Omega \biggr ]\cr
&-{2G\over 5c^5}\mu\left(\Ib5_{\bone\bone}r\dot r
+\Ib5_{\bone\btwo}r^2\Omega_{orb}\right ),    \cr
}\eqno(9.9)$$
where we have used equation (4.20) and $Mr_{cm}^2+M'r_{cm}'^2=\mu r^2$.

Using equation (4.21), the dissipative forces are then given by
$$\eqalign{
\cF_{a_1} &=-{2G\over 5c^5}\left({1\over 5}\kappa_nM\right )
\left(\Ib5_{\bone\bone}\cos^2\alpha+\Ib5_{\btwo\btwo}\sin^2\alpha
-\Ib5_{\bone\btwo}\sin 2\alpha\right)a_1,\cr
\cF_{a_2} &=-{2G\over 5c^5}\left({1\over 5}\kappa_nM\right )
\left(\Ib5_{\bone\bone}\sin^2\alpha+\Ib5_{\btwo\btwo}\cos^2\alpha
+\Ib5_{\bone\btwo}\sin 2\alpha\right)a_2,\cr
\cF_{a_3} &=-{2G\over 5c^5}\left({1\over 5}\kappa_nM\right )
\Ib5_{\bthr\bthr}a_3,\cr
\cF_\phi &=-{2G\over 5c^5}\left({1\over 5}\kappa_nM\right )
\left(\Ib5_{\bone\btwo}\cos 2\alpha
+(\Ib5_{\bone\bone}-\Ib5_{\btwo\btwo}){1\over 2}\sin 2\alpha\right)
(a_1^2-a_2^2),\cr
\cF_\psi &=0,\cr
\cF_r &=-{2G\over 5c^5}\Ib5_{\bone\bone}\mu r,\cr
\cF_\theta &=-{2G\over 5c^5}\Ib5_{\bone\btwo}\mu r^2.\cr
}\eqno(9.10)$$
Therefore, the dynamical equations for binaries, equations (5.11)--(5.17)
are  modified as
$$\eqalignno{
&\ddot a_1=\{\cdots\}-{2G\over 5c^5}\left[\Ib5_{\bone\bone}\cos^2\alpha
+\Ib5_{\btwo\btwo}\sin^2\alpha-\Ib5_{\bone\btwo}\sin 2\alpha\right]a_1,
&(9.11)\cr
&\ddot a_2=\{\cdots\}-{2G\over 5c^5}\left[\Ib5_{\bone\bone}\sin^2\alpha
+\Ib5_{\btwo\btwo}\cos^2\alpha+\Ib5_{\bone\btwo}\sin 2\alpha\right]a_2,
&(9.12)\cr
&\ddot a_3=\{\cdots\}-{2G\over 5c^5}\Ib5_{\bthr\bthr}a_3, &(9.13)\cr
&{d\over dt}\left(a_1\Omega-a_2\Lambda\right)=\{\cdots\}
-{2G\over 5c^5}\left[\Ib5_{\bone\btwo}\cos 2\alpha+
{1\over 2}(\Ib5_{\bone\bone}-\Ib5_{\btwo\btwo})\sin 2\alpha\right]a_1,
&(9.14)\cr
&{d\over dt}\left(-a_2\Omega+a_1\Lambda\right)=\{\cdots\}
-{2G\over 5c^5}\left[\Ib5_{\bone\btwo}\cos 2\alpha+
{1\over 2}(\Ib5_{\bone\bone}-\Ib5_{\btwo\btwo})\sin 2\alpha\right]a_2,
&(9.15)\cr
&\ddot r=\{\cdots\}-{2G\over 5c^5}\Ib5_{\bone\bone}r, &(9.16)\cr
&\ddot \theta=\{\cdots\}-{2G\over 5c^5}\Ib5_{\bone\btwo},  &(9.17)\cr
}$$
Since $\Ib5_{\bone\bone}+\Ib5_{\btwo\btwo}+\Ib5_{\bthr\bthr}=0$,
the expression for $2P_c/\rho_c$
is not affected by the presence of gravitational radiation reaction, i.e.,
equations~(2.30) and~(2.34) still apply. Finally, equations~(5.18) and~(5.19)
become
$$\eqalign{
\dot\Omega &=\left({a_2\over a_1}-{a_1\over a_2}\right)^{-1}
\left[\{\cdots\}+{2G\over 5c^5}\left(\Ib5_{\bone\btwo}\cos 2\alpha
+{1\over 2}(\Ib5_{\bone\bone}-\Ib5_{\btwo\btwo})\sin
2\alpha\right)\left({a_1\over a_2}
+{a_2\over a_1}\right)\right],\cr
\dot\Lambda &=\left({a_2\over a_1}-{a_1\over a_2}\right)^{-1}
\left[\{\cdots\}+{4G\over 5c^5}\left(\Ib5_{\bone\btwo}\cos 2\alpha
+{1\over 2}(\Ib5_{\bone\bone}-\Ib5_{\btwo\btwo})\sin 2\alpha\right)\right].\cr
}\eqno(9.18)$$
Note that the equation ${\cal F}_\psi=0$ again guarantees that the
fluid circulation defined by equation~(2.13) is conserved.

\bigskip
\centerline{\bf 9.2 Orbital Decay of Roche-Riemann Binaries}
\medskip

With the dynamical equations derived in \S 9.1 and the expressions
for $\Ib5_{\bi\bj}$ derived in Appendix~A, we can now
calculate the orbital evolution of a general Roche-Riemann
binary driven by gravitational radiation reaction.
In particular, we can study the dynamical behavior of
a coalescing neutron-star-black-hole pair prior to final merging,
at least when general relativistic effects are not too important
(i.e., when $r_{GR}\sim 6G(M+M')/c^2< (1+M'/M)^{1/3}R_o$; see LRS3, \S5).

At large separation, the orbital decay is secular,
and the binary evolves along a Roche-Riemann equilibrium sequence
with constant $\cC$ (since gravitational radiation reaction forces
conserve $\cC$). This quasi-static evolution has been discussed extensively in
LRS3. At smaller $r$, when the dynamical stability limit is
approached, the dynamical equations must be used.
In Figure~12 we show an example of
such a dynamical calculation for $p=1$, $n=1$, and $R_o=5GM/c^2$ (a typical
value for neutron stars; see LRS3). The fluid viscosity is assumed to be zero.
At $t=0$, we construct an equilibrium Roche-Riemann configuration
$\cC=0$ and $r/a_1=5$.
Initially, the binary closely follows the equilibrium constant$-\cC$
sequence. As the dynamical instability develops, both the
radial velocity and the dynamical tidal lag increase considerably.
Thereafter the two stars merge hydrodynamically in just a few orbits.
This qualitative behavior has already been observed in the simplified
calculations we presented in LRS2 and LRS3.
The development of a large lag angle is
not surprising. It arises because of the  finite
time necessary for the star to adjust its structure
to the rapidly changing tidal potential (cf.\ Lai 1994a).

The small but finite value of $\alpha$ observed in Figure~12 in the limit
of large orbital separation can be calculated as follows.
Including gravitational radiation reaction, the rate of change of
the spin $J_s$ is given by
$${dJ_s\over dt}
={3GM'\over 2r^3}\sin 2\alpha (I_{11}-I_{22})+\cF_\phi,\eqno(9.19)$$
(see eq.~[5.8]), where $\cF_\phi$ is given by equation (9.10).
Using the expressions derived in Appendix~A,
we see that the dominant contribution to $\cF_\phi$ comes from the terms
proportional to $\Ib5_{\bone\btwo}\simeq 16\Omega_{orb}^5\mu r^2$.
Thus
$${dJ_s\over dt}\simeq
{3GM'\over 2r^3}\sin 2\alpha (I_{11}-I_{22})
-{2G\over 5c^5}\left({1\over 5}\kappa_nM\right )
\left(\Ib5_{\bone\btwo}\cos 2\alpha\right)(a_1^2-a_2^2).\eqno(9.20)$$
As in \S 8.2, consider the limit of large $r$, so that $J_s\simeq -\cC$.
Since $d\cC/dt=0$ (cf.~eq.~[9.10]), we have $dJ_s/dt\simeq 0$.
Equation~(9.20) then gives
$$\tan 2\alpha\simeq {64\over 15}{\Omega_{orb}^5 G \mu r^2\over c^5M'}
\simeq {64\over 15}\left({G M\over rc^2}\right)^{5/2}
\left(1+{M'\over M}\right)^{3/2}.\eqno(9.21)$$
Note that this result is independent of the equation of state (the polytropic
index $n$ does not appear).
In Figure~13, we show the numerical results for $\alpha$ as a function of
$r$. We see that the value of $\alpha$ is indeed
given by equation~(9.21) when $r/R_o\gg1$. However, at smaller separation,
for $r/R_o\lo4$, the lag angle can become considerably larger
(by as much as an order of magnitude) than predicted by expression~(9.21).
This is  a result of the dynamical instability.

\bigskip
\bigskip
\centerline{\bf ACKNOWLEDGEMENTS}
\medskip

This work has been supported in part
by NSF Grant AST 91-19475 and NASA Grant NAGW-2364 to Cornell University.
Partial support was also provided by a Hubble Fellowship to F.~A.~R.
funded by NASA through Grant HF-1037.01-92A
from the Space Telescope Science Institute, which is operated by the
Association of Universities for Research in Astronomy, Inc.,
under contract NAS5-26555. F.~A.~R. also acknowledges the hospitality
of the ITP at UC Santa Barbara.

\vfill\eject
\bigskip
\centerline{\bf APPENDIX A: Evaluation of $\Ib5_{\bi\bj}$}
\medskip

Consider a triaxial body with its diagonal moment of inertia tensor
as defined in the body frame given by $I_{ij}$ (eq.\ [4.20]).
We now calculate $\Ib5_{\bi\bj}$ (cf.\ eq.~[9.2]), the components of
${\bf \Ibar}^{(5)}$ projected onto a ``projection frame.''
To define this frame, consider Figure~3, but for now neglect $M'$
by focusing on $M$: the body coordinates are $x_i$ and the projection
coordinates are $x_{\bi}$, but for now center the projection frame on O.
In the special case of a single star, when the projection frame coincides
with the body frame, the procedure for calculating $\Ib5_{\bi\bj}$
has been provided by Miller (1974). Below we generalize this procedure
to the case where $\{x_i\}$ and $\{x_\bi\}$ are different, as we have
for binaries (see \S 9).

The coordinates $\{x_i\}$ and $\{x_\bi\}$ are related
to the inertial coordinates by
$\{X_i\}$ by
$$x_i=T_{i\alpha}(\phi)X_\alpha,~~~~x_\bi=T_{\bi\alpha}(\theta)X_\alpha,
\eqno(A1)$$
where
$${\bf T}(\phi)=
\left(\matrix{\cos\phi &\sin\phi & 0\cr
		-\sin\phi &\cos\phi & 0\cr
		0 & 0 & 1\cr}\right),
\eqno(A2)$$
and similarly for ${\bf T}(\theta)$.
The components of the reduced quadrupole tensor in the inertial
frame are
$$\Ibar^{(in)}_{\alpha\beta}=T^{\dag}_{\alpha k}(\phi)T^{\dag}_{\beta
l}(\phi)\Ibar_{kl}.\eqno(A3)$$
Thus we have
$$\eqalign{
\Ib5_{\bi\bj} &=T_{\bi\alpha}(\theta)T_{\bj\beta}(\theta)
{d^5\over dt^5}[T^{\dag}_{\alpha k}(\phi)T^{\dag}_{\beta l}(\phi)\Ibar_{kl}]\cr
&=\sum_{m=0}^5 C_m^5\left({d^{5-m}\over dt^{5-m}}\Ibar_{kl}\right)
T_{\bi\alpha}(\theta)T_{\bj\beta}(\theta){d^m\over dt^m}
[T^{\dag}_{\alpha k}(\phi)T^{\dag}_{\beta l}(\phi)]\cr
&=\sum_{m=0}^5 C_m^5\left({d^{5-m}\over dt^{5-m}}\Ibar_{kl}\right)
\sum_{p=0}^m C^m_p R^p_{\bi k}R^{m-p}_{\bj l},\cr
}\eqno(A4)$$
where the constants $C_m^5$ are binomial coefficients, and
$$R^p_{\bi k}=T_{\bi\alpha}(\theta){d^p\over dt^p}T^{\dag}_{\alpha k}(\phi).
\eqno(A5)$$

In general, equations (A4)--(A5) are complicated to evaluate.
In the quasi-static limit, when $|da_i/dt|\ll|\Omega a_i|$,
simple expressions for $\Ib5_{ij}$ can be
derived. To lowest order in $da_i/dt$, we have
$$\Ib5_{\bi\bj}\simeq
\sum_k I_{kk}\sum_{p=0}^5C^5_p R^p_{\bi k}R^{5-p}_{\bj k}
+5\sum_k\dot I_{kk}\sum_{p=0}^4C^4_p R^p_{\bi k}R^{5-p}_{\bj k},\eqno(A6)$$
where we have used
$$\Ibar^{(in)}_{\alpha\beta}=T^{\dag}_{\alpha k}(\phi)T^{\dag}_{\beta l}(\phi)
I_{kl}-{1\over 3}I_{kk}\delta_{\alpha\beta},\eqno(A7)$$
and $dI_{kk}/dt\simeq 0$ in the quasi-static approximation.

The matrix $R^p$ can be now evaluated, keeping terms up to order $\dot\Omega$.
After some algebra, we obtain
$$\eqalign{
\biggl[\Ib5_{\bi\bj}\biggr]= &
16\Omega^5(I_{11}-I_{22})
\left(\matrix{\sin 2\alpha & \cos 2\alpha & 0\cr
		\cos 2\alpha & -\sin 2\alpha & 0\cr
		0 & 0 & 0\cr}\right)\cr
& +40\Omega^3[\Omega(\dot I_{11}-\dot I_{22})+2\dot\Omega (I_{11}-I_{22})]
\left(\matrix{\cos 2\alpha & -\sin 2\alpha & 0\cr
		-\sin 2\alpha & -\cos 2\alpha & 0\cr
		0 & 0 & 0\cr}\right).\cr
}\eqno(A8)$$

We can now write down the components $\Ib5_{\bi\bj}$ that appear in our
dynamical
equations. For the single star case, the projection frame coincides with
the body frame. Setting $\alpha=0$ in equation (A8), we see that
the only nonzero components are
$$\eqalign{
\Ib5_{11}=-\Ib5_{22}&=40\Omega^3[\Omega(\dot I_{11}-\dot I_{22})+2\dot\Omega
(I_{11}-I_{22})],\cr
\Ib5_{12}=\Ib5_{21}&=16\Omega^5(I_{11}-I_{22}),\cr
}\eqno(A9)$$
where the $I_{ii}$ are defined in equation (4.20).

To evaluate $\Ib5_{\bi\bj}$ for a Roche-Riemann binary, we now move the
origin of the projection coordinates $\{x_{\bi}\}$ back to the system CM
as in Figure~3 and equation~(9.1). But then we can decompose the total moment
of inertia into two contributions: the orbital part due to two point masses
$M$ and $M'$, and the fluid part due to the ellipsoid $M$. The contribution
from
the ellipsoid is given by equation~(A8). The resulting nonzero components
of $\Ib5_{\bi\bj}$ are
$$\eqalign{
\Ib5_{\bone\bone}=-\Ib5_{\btwo\btwo}
&= 16\Omega^5(I_{11}-I_{22})\sin 2\alpha+40\Omega_{orb}^3
[2\Omega_{orb}\mu r\dot r+2\dot\Omega_{orb} \mu r^2]\cr
&~~~~~~~~~~~~~~~~~~+40\Omega^3[\Omega(\dot I_{11}
-\dot I_{22})+2\dot\Omega(I_{11}-I_{22})]\cos 2\alpha,\cr
\Ib5_{\bone\btwo}=\Ib5_{\btwo\bone}
&=16\Omega_{orb}^5\mu r^2+16\Omega^5(I_{11}-I_{22})\cos 2\alpha\cr
&~~~~~~~~~~~~~~~~~~-40\Omega^3[\Omega
(\dot I_{11}-\dot I_{22})+2\dot\Omega (I_{11}-I_{22})]\sin 2\alpha.\cr
}\eqno(A10)$$

These expressions obviously do not apply to
certain dynamical situations, such as
the hyperbolic fly-by of a star past a black-hole. In such a case
the variables $r$ and $a_i$ change rapidly and higher-order time
derivatives must be retained. For two point masses, the expressions
for $\Ib5_{\bi\bj}$ can be evaluated analytically with repeated use
of the equation of motion.

\vfill\eject
\centerline{\bf REFERENCES}
\medskip
\def\bysame{\hbox to 50pt{\leaders\hrule height 2.4pt depth -2pt\hfill .\ }}
\def\hi{\noindent \hangindent=2.5em}

\hi{
Aizenman, M.L. 1968, ApJ, 153, 511}

\hi{
Boss, A.P., Cameron, A.G.W., \& Benz, W. 1991, Icarus, 92, 165}

\hi{
Carter, B., \& Luminet, J.P. 1985, MNRAS, 212, 23}

\hi{
Chandrasekhar, S. 1961, Hydrodynamics and Hydromagnetic
Stability (Oxford Univ. Press: Oxford)}

\hi{
\bysame, 1969, Ellipsoidal Figures of Equilibrium
(New Haven: Yale University Press) (Ch69)}

\hi{
\bysame 1970, ApJ, 161, 561}

\hi{
Cox, J.P. 1980, Theory of Stellar Pulsation (Princeton University Press)}

\hi{
Detweiler, S.L., \& Lindblom, L. 1977, ApJ, 213, 193}

\hi{
Evans, C.R., \& Kochanek, C.S. 1989, ApJ, 346, L13}

\hi{
Fabian, A.C., Pringle, J.E., \& Rees, M.J. 1975, MNRAS, 172, 15P}

\hi{
Goldreich, P., \& Peale, S.J. 1968, ARAA, 6, 287}

\hi{
Goldstein, H. 1980, Classical Mechanics (Reading: Addison-Wesley)}

\hi{
Ipser, J., \& Managan, M. 1981, ApJ, 256, 145

\hi{
Kochanek C.S. 1992a, ApJ, 385, 604}

\hi{
\bysame 1992b, ApJ, 398, 234}

\hi{
Kosovichev, A.G., \& Novikov, I.D. 1992, MNRAS, 258, 715}

\hi{
Laguna, P., Miller, W.A., Zurek, W.H., \& Davies, M.B. 1993, ApJL, 410, L83}

\hi{
Lai, D. 1994a, MNRAS, submitted}

\hi{
\bysame 1994b, Ph.D. thesis, Cornell University}

\hi{
Lai, D., Rasio, F.A., \& Shapiro, S.L. 1993a, ApJS, 88, 205 (LRS1)}

\hi{
\bysame 1993b, ApJL, 406, L63 (LRS2)}

\hi{
\bysame 1993c, ApJ, 412, 593}

\hi{
\bysame 1994a, ApJ, 420, 811 (LRS3)}

\hi{
\bysame 1994b, ApJ, 423, 344 (LRS4)}

\hi{
\bysame 1994c, in preparation}

\hi{
Lai, D., \& Shapiro, S.L. 1994, in preparation}

\hi{
Landau, L.D., \& Lifshitz, E.M. 1987, Fluid Mechanics, 2nd
Ed.\ (Oxford: Pergamon Press)}

\hi{
Lebovitz, N.R. 1966, ApJ, 145, 878}

\hi{
Lee, H.M., \& Ostriker, J.P. 1986, ApJ, 310, 176}

\hi{
Lindblom, L, \& Detweiler 1977, ApJ, 211, 565}

\hi{
Luminet, J.P., \& Carter, B. 1986, ApJS, 61, 219}

\hi{
McMillan, S.L.W., McDermott, P.N., \& Taam, R.E. 1987, ApJ, 318, 261}

\hi{
Miller, B.D. 1974, ApJ, 187, 609}

\hi{
Misner, C.M., Thorne, K.S., \& Wheeler, J.A. 1970,
{\it Gravitation} (New York: Freeman)}

\hi{
Novikov, I.D., Pethick, C.J., \& Polnarev, A.G. 1992, MNRAS, 255, 276}

\hi{
Nduka, A. 1971, ApJ, 170, 131}

\hi{
Press, W.H., \& Teukolsky, S.A. 1973, ApJ, 181, 513}

\hi{
\bysame 1977, ApJ, 213, 183 (PT)}

\hi{
Press, W.~H., Teukolsky, S.~A., Vetterling, W.~T, \& Flannery, B.~P. 1992,
Numerical Recipes: The Art of Scientific Computing, 2nd Ed.
(Cambridge: Cambridge Univ.\ Press)}

\hi{
Rasio, F. A. 1993, PASP, 105, 973}

\hi{
Rasio, F. A., \& Shapiro, S.L. 1991, ApJ, 377, 559}

\hi{
Rasio, F. A., \& Shapiro, S.L. 1994, ApJ, in press (September 1)}

\hi{
Rees, M.J. 1988, Nature, 333, 523}

\hi{
Roberts, P.H., \& Stewartson, K. 1963, ApJ, 137, 777}

\hi{
Rossner, L.F. 1967, ApJ, 149, 145}

\hi{
Shapiro, S.L. 1979, in Sources of Gravitational Radiation, ed. L. Smarr
(Cambridge Univ. Press)}


\hi{
Sridhar, S., \& Tremaine, S. 1992, Icarus, 95, 86}

\hi{
Tassoul, J.-L. 1970, ApJ, 160, 1031}

\hi{
\bysame 1978, Theory of Rotating Stars
(Princeton: Princeton University Press)}


\hi{
Zahn, J.P. 1970, A\&A, 4, 452}

\vfill\eject
\midinsert
\leftskip 0.25in
\rightskip 0.25in
\newdimen\digitwidth
\setbox0=\hbox{\rm0}
\digitwidth=\wd0
\catcode`?=\active
\def?{\kern\digitwidth}
$$\vbox{
\tabskip=2em plus3em minus1.5em
\halign to3in{
\hfil#\hfil& \hfil#\hfil& \hfil#\hfil \cr
\multispan3\hfil {\bf TABLE 1} \hfil\cr
\noalign{\smallskip}
\multispan3\hfil {Radial Pulsations of Polytropes$^a$} \hfil\cr
\noalign{\medskip \hrule \vskip1pt \hrule \smallskip}
$n$ & ${\hat\sigma}_o^2$ & $q_n^{-1}$ \cr
\noalign{\smallskip \hrule \smallskip}
0.&	1.&	1. \cr
0.5&	1.355&	1.364 \cr
1.0&	1.877&	1.913 \cr
1.5&	2.706&	2.793 \cr
2.0&	4.137&	4.305 \cr
2.5&	6.909&	7.155 \cr
3.0&	13.27&	13.27 \cr
\noalign{\smallskip \hrule}
}
}$$
\hskip 0.3in NOTE: $^a$ Here ${\hat\sigma}_o^2\equiv
\sigma_o^2/[(3\Gamma-4)GM/R_o^3]$ and $\Gamma=\Gamma_1=1+1/n$.
\endinsert

\midinsert
\leftskip 0.25in
\rightskip 0.25in
\newdimen\digitwidth
\setbox0=\hbox{\rm0}
\digitwidth=\wd0
\catcode`?=\active
\def?{\kern\digitwidth}
$$\vbox{
\tabskip=2em plus3em minus1.5em
\halign to3in{
\hfil#\hfil& \hfil#\hfil& \hfil#\hfil \cr
\multispan3\hfil {\bf TABLE 2} \hfil\cr
\noalign{\smallskip}
\multispan3\hfil {Disruption Limits and Roche-Riemann Limits$^a$}\hfil\cr
\noalign{\medskip \hrule \vskip1pt \hrule \smallskip}
$n$ & $\eta_{dis}$ & $\eta_{rr}$ \cr
\noalign{\smallskip \hrule \smallskip}
0.&     2.21 & 3.806 \cr
0.5&	2.08 & 3.427\cr
1.0&	1.96 & 3.054\cr
1.5&	1.84 & 2.698\cr
2.0&	1.71 & 2.368\cr
2.5&	1.584& 2.117\cr
\noalign{\smallskip \hrule}
}
}$$
\hskip 0.3in NOTE: $^a$ Here $\eta_{dis}$ is the disruption
limit for encounters between
a star and a massive body, $\eta_{rr}$ is the Roche-Riemann limit for
equilibrium binaries in circular orbits.
\endinsert

\vfil\eject
\centerline{\bf FIGURE CAPTIONS}
\vskip 0.3truecm

\noindent
{\bf FIG.~1}.---
Secular evolution tracks of a Riemann-S ellipsoid with $n=1$ driven by
gravitational radiation reaction. The energy of an ellipsoid as a function of
the axis ratio $a_2/a_1$ is shown along various equilibrium sequences.
Dedekind-like sequences ($|\Lambda|>|\Omega|$) are shown on the left,
Jacobi-like ($|\Omega|>|\Lambda|$) on the right.
The thick solid line corresponds to the Dedekind and Jacobi
sequences, while the other lines correspond to constant-$\cC$ sequences:
$\bar \cC=\cC/(GM^3R_o)^{1/2}=-0.48$ (dotted line),
$\bar\cC=-0.4$ (dashed line), $\bar\cC=-0.32$ (long dashed line)
and $\bar\cC=-0.25$ (dotted-dashed line).
The solid round dot marks the point of bifurcation.
This figure can also be applied to pure viscous evolution, in which case
the curves for Jacobi-like sequences and Dedekind-like sequences are switched,
and $\cC$ is replaced by $J=-\cC=$constant.

\medskip\noindent
{\bf FIG.~2}.---
Energy as a function of angular momentum for
Maclaurin (solid lines), Jacobi (dotted ines) and
Dedekind (dashed lines) equilibrium sequences, for $n=0,~1,~2$.

\medskip\noindent
{\bf FIG.~3}.---
A sketch of the coordinate systems used for binaries.

\medskip\noindent
{\bf FIG.~4}.---
Evolution of the binary separation $r$, the axes $a_i$, and the lag angle
$\alpha$ for Roche-Riemann binaries, with
$p=M/M'=1$ and $n=1$. The initial configurations are in equilibrium,
and corotating.
The left panels show the evolution of a dynamically unstable binary
($r/a_1=1.7, r/R_o=2.406$), while the right panels show that of
a stable binary ($r/a_1=1.8, r/R_o=2.427$). Both equilibrium
configurations are perturbed by imposing a small radial velocity.

\medskip\noindent
{\bf FIG.~5}.---
Evolution of the axes, energy, and angular momentum of a star with $n=1.5$
 during a parabolic encounter with a massive ($M'\gg M$) body. The
parameter $\eta=2.8$ (eq.\ [7.1]).
The solid line in the middle shows the total (conserved) energy in the system.

\medskip\noindent
{\bf FIG.~6}.---
Energy transfer and angular momentum transfer to the star during its parabolic
encounter with a massive body. The heavy lines are the results obtained by
integrating numerically our equations for a compressible ellipsoid,
while the lighter lines show the results of linear perturbation theory
(eqs.~[7.5] and [7.8]). The solid lines are for $n=0$,
the short-dashed lines for $n=1.5$, and the long-dashed lines for $n=2.5$.
In the linear theory, all modes are calculated assuming $\Gamma_1=\Gamma$,
except in the case of $n=2.5$, where we have also
included curves (dotted lines) showing the results when $\Gamma_1=5/3$.

\medskip\noindent
{\bf FIG.~7}.---
Same as Fig.~5, but for $\eta=1.85$ (left) and $\eta=1.83$ (right).
The dashed lines in the middle  show the orbital energy.

\medskip\noindent
{\bf FIG.~8}.---
Same as Fig.~7, but here the viscosity is nonzero,
with $\nu=0.01(GMR_o)^{1/2}$.
On the left $\eta=1.73$ and on the right $\eta=1.71$.

\medskip\noindent
{\bf FIG.~9}.---
The disruption limit $\eta_{dis}$ for the parabolic encounter of a star with a
massive body, as a function of the fluid viscosity $\nu$.
The viscosity is assumed to be constant during the
encounter. The three curves correspond to $n=0.1$ (solid line),
$n=1.5$ (short-dashed line) and $n=2.5$ (long-dashed line).

\medskip\noindent
{\bf FIG.~10}.---
The total energy $E_{eq}$ of several equilibrium Roche-Riemann sequences
for $p=M/M'=1$ and $n=0$, as a function of the binary separation $r$.
The solid line is for the corotating sequence ($\Lambda=0$),
the dotted line for a sequence with constant
${\bar J}=J/(GM^3R_o)^{1/2}=1.4>{\bar J}_{crit}=1.375$, and the dashed line
for a sequence with constant ${\bar J}=1.3<{\bar J}_{crit}$.

\medskip\noindent
{\bf FIG.~11}.---
The evolution of a Roche-Riemann binary driven by viscosity. Here $p=1$,
$n=1$, and $\nu=0.01(GMR_o)^{1/2}$. The initial configuration is an equilibrium
Roche-Riemann binary with $f_R=-4$ and $r/a_1=3$, corresponding to
$J=1.220(GM^3R_o)^{1/2}$.
Here $r$ is the binary separation,
$a_i$ are the ellipsoid's axes, $\Lambda$ measures the degree of
nonsynchronization
($\Lambda\simeq \Omega-\Omega_s$), $\alpha$ is the tidal lag angle, and
$E$ is the total energy of the system.

\medskip\noindent
{\bf FIG.~12}.---
The evolution of a Roche-Riemann binary driven by gravitational radiation.
Here $p=1$, $n=1$, $\nu=0$ and $R_o=5GM/c^2$.
The initial configuration is an equilibrium Roche-Riemann binary
with $f_R=-2$ and $r/a_1=5$. All quantities are defined as in
Fig.~11, and $v_r=\dot r$ is the radial velocity.
The axes for an equilibrium irrotational Roche-Riemann
sequence are shown as lighter lines, while the dynamical values are
shown as thicker lines. The total energy for an equilibrium sequence
is also shown as a dotted line.

\medskip\noindent
{\bf FIG.~13}.---
The tidal lag angle as a function of binary separation $r$. Here $p=1$, $n=0.5$
and $R_o=5GM/c^2$. The solid line is from our dynamical calculation,
the dotted line is an analytic expression for large $r$ (eq.~[9.21]).

\end